\input harvmac.tex

\input epsf.tex

\def\figin{\epsfcheck\figin}\def\figins{\epsfcheck\figins}
\def\epsfcheck{\ifx\epsfbox\UnDeFiNeD
\message{(NO epsf.tex, FIGURES WILL BE IGNORED)}
\gdef\figin##1{\vskip2in}\gdef\figins##1{\hskip.5in}
\else\message{(FIGURES WILL BE INCLUDED)}%
\gdef\figin##1{##1}\gdef\figins##1{##1}\fi}
\def\DefWarn#1{}
\def\figinsert{\goodbreak\midinsert}
\def\ifig#1#2#3{\DefWarn#1\xdef#1{figure~\the\figno}
\writedef{#1\leftbracket figure\noexpand~\the\figno}%
\figinsert\figin{\centerline{#3}}\medskip\centerline{\vbox{\baselineskip12pt
\advance\hsize by -1truein\noindent\footnotefont{\bf
Figure~\the\figno:} #2}}
\bigskip\endinsert\global\advance\figno by1}



\lref\integrabilityft{
  J.~A.~Minahan and K.~Zarembo,
  ``The Bethe ansatz for superconformal Chern-Simons,''
  arXiv:0806.3951 [hep-th];
  D.~Gaiotto, S.~Giombi and X.~Yin,
  ``Spin Chains in N=6 Superconformal Chern-Simons-Matter Theory,''
  arXiv:0806.4589 [hep-th];
  D.~Bak and S.~J.~Rey,
  ``Integrable Spin Chain in Superconformal Chern-Simons Theory,''
  arXiv:0807.2063 [hep-th].
  }
\lref\integrabilityst{
 G.~Arutyunov and S.~Frolov,
  ``Superstrings on $AdS_4$ x $CP^3$ as a Coset Sigma-model,''
  arXiv:0806.4940 [hep-th];
  B.~Stefanski,
  ``Green-Schwarz action for Type IIA strings on $AdS_4\times CP^3$,''
  arXiv:0806.4948 [hep-th];
  G.~Grignani, T.~Harmark and M.~Orselli,
  ``The SU(2) x SU(2) sector in the string dual of N=6 superconformal
  Chern-Simons theory,''
  arXiv:0806.4959 [hep-th];
  G.~Grignani, T.~Harmark, M.~Orselli and G.~W.~Semenoff,
  ``Finite size Giant Magnons in the string dual of N=6 superconformal
  Chern-Simons theory,''
  arXiv:0807.0205 [hep-th];
  N.~Gromov and P.~Vieira,
  ``The AdS4/CFT3 algebraic curve,''
  arXiv:0807.0437 [hep-th];
  C.~Ahn and P.~Bozhilov,
  ``Finite-size effects of Membranes on $AdS_4\times S_7$,''
  arXiv:0807.0566 [hep-th];
  N.~Gromov and P.~Vieira,
  ``The all loop AdS4/CFT3 Bethe ansatz,''
  arXiv:0807.0777 [hep-th];
  B.~Chen and J.~B.~Wu,
  ``Semi-classical strings in $AdS_4$*$CP^3$,''
  arXiv:0807.0802 [hep-th];
  D.~Astolfi, V.~G.~M.~Puletti, G.~Grignani, T.~Harmark and M.~Orselli,
  ``Finite-size corrections in the SU(2) x SU(2) sector of type IIA string
  theory on $AdS_4$ x $CP^3$,''
  arXiv:0807.1527 [hep-th];
  C.~Ahn and R.~I.~Nepomechie,
  ``N=6 super Chern-Simons theory S-matrix and all-loop Bethe ansatz
  equations,''
  arXiv:0807.1924 [hep-th];
  B.~H.~Lee, K.~L.~Panigrahi and C.~Park,
  ``Spiky Strings on AdS$_4 \times {\bf CP}^3$,''
  arXiv:0807.2559 [hep-th];
  C.~Ahn, P.~Bozhilov and R.~C.~Rashkov,
  ``Neumann-Rosochatius integrable system for strings on $AdS_4\times CP^3$,''
  arXiv:0807.3134 [hep-th];
  T.~McLoughlin and R.~Roiban,
  ``Spinning strings at one-loop in $AdS_4 \times P^3$,''
  arXiv:0807.3965 [hep-th];
  L.~F.~Alday, G.~Arutyunov and D.~Bykov,
  ``Semiclassical Quantization of Spinning Strings in $AdS_4 \times CP^3$,''
  arXiv:0807.4400 [hep-th];
  C.~Krishnan,
  ``$AdS_4/CFT_3$ at one loop,''
  arXiv:0807.4561 [hep-th].
}

\lref\ArmoniKR{
  A.~Armoni and A.~Naqvi,
  ``A Non-Supersymmetric Large-N 3D CFT And Its Gravity Dual,''
  arXiv:0806.4068 [hep-th].
}

\lref\HananyQC{
  A.~Hanany, N.~Mekareeya and A.~Zaffaroni,
  ``Partition Functions for Membrane Theories,''
  arXiv:0806.4212 [hep-th].
}

\lref\wittenbaryon{
  E.~Witten,
  ``Baryons and branes in anti de Sitter space,''
  JHEP {\bf 9807}, 006 (1998)
  [arXiv:hep-th/9805112].
}

\lref\kao{
  B.~M.~Zupnik and D.~V.~Khetselius,
  ``Three-dimensional extended supersymmetry in harmonic superspace,''
  Sov.\ J.\ Nucl.\ Phys.\  {\bf 47}, 730 (1988)
  [Yad.\ Fiz.\  {\bf 47}, 1147 (1988)];
  H.~C.~Kao,
  ``Selfdual Yang-Mills Chern-Simons Higgs systems with an N=3 extended
  supersymmetry,''
  Phys.\ Rev.\  D {\bf 50}, 2881 (1994);
  }
 \lref\cscoefficient{ H.~C.~Kao, K.~M.~Lee and T.~Lee,
  ``The Chern-Simons coefficient in supersymmetric Yang-Mills Chern-Simons
  theories,''
  Phys.\ Lett.\  B {\bf 373}, 94 (1996)
  [arXiv:hep-th/9506170].
}

\lref\klebanovtseytlin{
  I.~R.~Klebanov and A.~A.~Tseytlin,
  ``Entropy of Near-Extremal Black p-branes,''
  Nucl.\ Phys.\  B {\bf 475}, 164 (1996)
  [arXiv:hep-th/9604089].
}

\lref\MaldacenaIM{
  J.~M.~Maldacena,
  ``Wilson loops in large N field theories,''
  Phys.\ Rev.\ Lett.\  {\bf 80}, 4859 (1998)
  [arXiv:hep-th/9803002];
  S.~J.~Rey and J.~T.~Yee,
  ``Macroscopic strings as heavy quarks in large N gauge theory and  anti-de
  Sitter supergravity,''
  Eur.\ Phys.\ J.\  C {\bf 22}, 379 (2001)
  [arXiv:hep-th/9803001].
}

\lref\MaldacenaRE{
  J.~M.~Maldacena,
  ``The large N limit of superconformal field theories and supergravity,''
  Adv.\ Theor.\ Math.\ Phys.\  {\bf 2}, 231 (1998)
  [Int.\ J.\ Theor.\ Phys.\  {\bf 38}, 1113 (1999)]
  [arXiv:hep-th/9711200].
}

\lref\wittencp{
  Y.~Y.~Goldschmidt and E.~Witten,
  ``Conservation Laws In Some Two-Dimensional Models,''
  Phys.\ Lett.\  B {\bf 91}, 392 (1980).
}

\lref\abdalla{
  M.~Gomes, E.~Abdalla and M.~C.~B.~Abdalla,
  ``On The Nonlocal Charge Of The $CP^{N-1}$ Model And Its Supersymmetric
  Extension To All Orders,''
  Phys.\ Rev.\  D {\bf 27}, 825 (1983).
}

\lref\gaiottoyin{
  D.~Gaiotto and X.~Yin,
  ``Notes on superconformal Chern-Simons-matter theories,''
  JHEP {\bf 0708}, 056 (2007)
  [arXiv:0704.3740 [hep-th]].
}

\lref\wittenthree{
  E.~Witten,
  ``Supersymmetric index of three-dimensional gauge theory,''
  arXiv:hep-th/9903005.
}

\lref\BiranIY{
  B.~Biran, A.~Casher, F.~Englert, M.~Rooman and P.~Spindel,
  ``The Fluctuating Seven Sphere In Eleven-Dimensional Supergravity,''
  Phys.\ Lett.\  B {\bf 134}, 179 (1984);
  L.~Castellani, R.~D'Auria, P.~Fre, K.~Pilch and P.~van Nieuwenhuizen,
  ``The Bosonic Mass Formula For Freund-Rubin Solutions Of D = 11 Supergravity
  On General Coset Manifolds,''
  Class.\ Quant.\ Grav.\  {\bf 1}, 339 (1984).
}

\lref\senteight{
  A.~Sen,
  ``Orbifolds of M theory and String Theory,''
  Mod.\ Phys.\ Lett.\  A {\bf 11}, 1339 (1996)
  [arXiv:hep-th/9603113].
}

\lref\kkmonopole{
  J.~P.~Gauntlett, G.~W.~Gibbons, G.~Papadopoulos and P.~K.~Townsend,
  ``Hyper-Kaehler manifolds and multiply intersecting branes,''
  Nucl.\ Phys.\  B {\bf 500}, 133 (1997)
  [arXiv:hep-th/9702202].
}

\lref\McGreevyCW{
  J.~McGreevy, L.~Susskind and N.~Toumbas,
  ``Invasion of the giant gravitons from anti-de Sitter space,''
  JHEP {\bf 0006}, 008 (2000)
  [arXiv:hep-th/0003075].
}

\lref\tHooftHY{
  G.~'t Hooft,
  ``On The Phase Transition Towards Permanent Quark Confinement,''
  Nucl.\ Phys.\  B {\bf 138}, 1 (1978).
}

\lref\AharonyRM{
  O.~Aharony, Y.~Oz and Z.~Yin,
  ``M theory on AdS(p) x S(11-p) and superconformal field theories,''
  Phys.\ Lett.\  B {\bf 430}, 87 (1998)
  [arXiv:hep-th/9803051];
  S.~Minwalla,
  ``Particles on AdS(4/7) and primary operators on M(2/5) brane
  worldvolumes,''
  JHEP {\bf 9810}, 002 (1998)
  [arXiv:hep-th/9803053];
  E.~Halyo,
  ``Supergravity on AdS(4/7) x S(7/4) and M branes,''
  JHEP {\bf 9804}, 011 (1998)
  [arXiv:hep-th/9803077].
}

\lref\emss{
  S.~Elitzur, G.~W.~Moore, A.~Schwimmer and N.~Seiberg,
  ``Remarks On The Canonical Quantization Of The Chern-Simons-Witten Theory,''
  Nucl.\ Phys.\  B {\bf 326}, 108 (1989).
}
\lref\bergman{
  T.~Kitao, K.~Ohta and N.~Ohta,
  ``Three-dimensional gauge dynamics from brane configurations with
  (p,q)-fivebrane,''
  Nucl.\ Phys.\  B {\bf 539}, 79 (1999)
  [arXiv:hep-th/9808111];
  O.~Bergman, A.~Hanany, A.~Karch and B.~Kol,
  ``Branes and supersymmetry breaking in 3D gauge theories,''
  JHEP {\bf 9910}, 036 (1999)
  [arXiv:hep-th/9908075];
    K.~Ohta,
  ``Supersymmetric index and s-rule for type IIB branes,''
  JHEP {\bf 9910}, 006 (1999)
  [arXiv:hep-th/9908120].
}
\lref\wittensl{
  E.~Witten,
  ``SL(2,Z) action on three-dimensional conformal field theories with Abelian
  symmetry,''
  arXiv:hep-th/0307041.
}
\lref\TomasielloEQ{
  A.~Tomasiello,
  ``New string vacua from twistor spaces,''
  arXiv:0712.1396 [hep-th].
}
\lref\wittenjones{
  E.~Witten,
  ``Quantum field theory and the Jones polynomial,''
  Commun.\ Math.\ Phys.\  {\bf 121}, 351 (1989).
}

\lref\becker{
  K.~Becker and M.~Becker,
  ``M theory on Eight-Manifolds,''
  Nucl.\ Phys.\  B {\bf 477}, 155 (1996)
  [arXiv:hep-th/9605053].
}

\lref\lamberttong{
  N.~Lambert and D.~Tong,
  ``Membranes on an Orbifold,''
  arXiv:0804.1114 [hep-th].
}

\lref\HalyoPN{
  E.~Halyo,
  ``Supergravity on AdS(5/4) x Hopf fibrations and conformal field  theories,''
  Mod.\ Phys.\ Lett.\  A {\bf 15}, 397 (2000)
  [arXiv:hep-th/9803193].
}

\lref\SethiZK{
  S.~Sethi,
  ``A relation between N = 8 gauge theories in three dimensions,''
  JHEP {\bf 9811}, 003 (1998)
  [arXiv:hep-th/9809162].
}

\lref\kapustin{
  A.~Kapustin,
  ``Wilson-'t Hooft operators in four-dimensional gauge theories and
  S-duality,''
  Phys.\ Rev.\  D {\bf 74}, 025005 (2006)
  [arXiv:hep-th/0501015].
  }

\lref\martelli{
  D.~Martelli and J.~Sparks,
  ``Dual giant gravitons in Sasaki-Einstein backgrounds,''
  Nucl.\ Phys.\  B {\bf 759}, 292 (2006)
  [arXiv:hep-th/0608060].
}

\lref\susywosusy{
  M.~J.~Duff, H.~Lu and C.~N.~Pope,
  ``Supersymmetry without supersymmetry,''
  Phys.\ Lett.\  B {\bf 409}, 136 (1997)
  [arXiv:hep-th/9704186].
}

\lref\gkp{
  S.~S.~Gubser, I.~R.~Klebanov and A.~M.~Polyakov,
  ``A semi-classical limit of the gauge/string correspondence,''
  Nucl.\ Phys.\  B {\bf 636}, 99 (2002)
  [arXiv:hep-th/0204051].
}

\lref\aldayjmtwo{
  L.~F.~Alday and J.~M.~Maldacena,
  ``Comments on operators with large spin,''
  JHEP {\bf 0711}, 019 (2007)
  [arXiv:0708.0672 [hep-th]].
}

\lref\WittenZW{
  E.~Witten,
  ``Anti-de Sitter space, thermal phase transition, and confinement in  gauge
  theories,''
  Adv.\ Theor.\ Math.\ Phys.\  {\bf 2}, 505 (1998)
  [arXiv:hep-th/9803131].
}

\lref\tHooftJZ{
  G.~'t Hooft,
  ``A planar diagram theory for strong interactions,''
  Nucl.\ Phys.\  B {\bf 72}, 461 (1974).
}


\lref\schwarz{
  J.~H.~Schwarz,
  ``Superconformal Chern-Simons theories,''
  JHEP {\bf 0411}, 078 (2004)
  [arXiv:hep-th/0411077].
}

\lref\GubserDE{
  S.~S.~Gubser, I.~R.~Klebanov and A.~W.~Peet,
  Phys.\ Rev.\  D {\bf 54}, 3915 (1996)
  [arXiv:hep-th/9602135].
}

\lref\baggerlambert{
  J.~Bagger and N.~Lambert,
  ``Comments On Multiple M2-branes,''
  JHEP {\bf 0802}, 105 (2008)
  [arXiv:0712.3738 [hep-th]];
  J.~Bagger and N.~Lambert,
  ``Gauge Symmetry and Supersymmetry of Multiple M2-Branes,''
  Phys.\ Rev.\  D {\bf 77}, 065008 (2008)
  [arXiv:0711.0955 [hep-th]];
  J.~Bagger and N.~Lambert,
  ``Modeling multiple M2's,''
  Phys.\ Rev.\  D {\bf 75}, 045020 (2007)
  [arXiv:hep-th/0611108].
}

\lref\markvr{
  M.~Van Raamsdonk,
  ``Comments on the Bagger-Lambert theory and multiple M2-branes,''
  arXiv:0803.3803 [hep-th].
}


\lref\KlebanovHH{
  I.~R.~Klebanov and E.~Witten,
  ``Superconformal field theory on threebranes at a Calabi-Yau  singularity,''
  Nucl.\ Phys.\  B {\bf 536}, 199 (1998)
  [arXiv:hep-th/9807080].
}

\lref\dmpv{
  J.~Distler, S.~Mukhi, C.~Papageorgakis and M.~Van Raamsdonk,
  ``M2-branes on M-folds,''
  JHEP {\bf 0805}, 038 (2008)
  [arXiv:0804.1256 [hep-th]].
}

\lref\ItzhakiRC{
  N.~Itzhaki,
  ``Anyons, 't Hooft loops and a generalized connection in three dimensions,''
  Phys.\ Rev.\  D {\bf 67}, 065008 (2003)
  [arXiv:hep-th/0211140].
}

\lref\lamberttong{
  N.~Lambert and D.~Tong,
  ``Membranes on an Orbifold,''
  arXiv:0804.1114 [hep-th].
}

\lref\IntriligatorEX{
  K.~A.~Intriligator and N.~Seiberg,
  ``Mirror symmetry in three dimensional gauge theories,''
  Phys.\ Lett.\  B {\bf 387}, 513 (1996)
  [arXiv:hep-th/9607207].
}

\lref\baggerlambert{
  J.~Bagger and N.~Lambert,
  ``Comments On Multiple M2-branes,''
  JHEP {\bf 0802}, 105 (2008)
  [arXiv:0712.3738 [hep-th]];
  J.~Bagger and N.~Lambert,
  ``Gauge Symmetry and Supersymmetry of Multiple M2-Branes,''
  Phys.\ Rev.\  D {\bf 77}, 065008 (2008)
  [arXiv:0711.0955 [hep-th]];
  J.~Bagger and N.~Lambert,
  ``Modeling multiple M2's,''
  Phys.\ Rev.\  D {\bf 75}, 045020 (2007)
  [arXiv:hep-th/0611108].
}

\lref\nonrenold{
  A.~N.~Kapustin and P.~I.~Pronin,
  ``Nonrenormalization theorem for gauge coupling in (2+1)-dimensions,''
  Mod.\ Phys.\ Lett.\  A {\bf 9}, 1925 (1994)
  [arXiv:hep-th/9401053].
 W.~Chen, G.~W.~Semenoff and Y.~S.~Wu,
  ``Two loop analysis of nonAbelian Chern-Simons theory,''
  Phys.\ Rev.\  D {\bf 46}, 5521 (1992)
  [arXiv:hep-th/9209005].
  }
  \lref\kapustinstrassler{
    A.~Kapustin and M.~J.~Strassler,
  ``On mirror symmetry in three dimensional Abelian gauge theories,''
  JHEP {\bf 9904}, 021 (1999)
  [arXiv:hep-th/9902033].
}

\lref\GustavssonVU{
  A.~Gustavsson,
  ``Algebraic structures on parallel M2-branes,''
  arXiv:0709.1260 [hep-th];
    A.~Gustavsson,
  ``Selfdual strings and loop space Nahm equations,''
  JHEP {\bf 0804}, 083 (2008)
  [arXiv:0802.3456 [hep-th]].
}

\lref\susywosusy{
  M.~J.~Duff, H.~Lu and C.~N.~Pope,
  ``Supersymmetry without supersymmetry,''
  Phys.\ Lett.\  B {\bf 409}, 136 (1997)
  [arXiv:hep-th/9704186].
}

\lref\dtwomtwo{
  S.~Mukhi and C.~Papageorgakis,
  ``M2 to D2,''
  arXiv:0803.3218 [hep-th].
}

\lref\nimadec{
  N.~Arkani-Hamed, A.~G.~Cohen, D.~B.~Kaplan, A.~Karch and L.~Motl,
  ``Deconstructing (2,0) and little string theories,''
  JHEP {\bf 0301}, 083 (2003)
  [arXiv:hep-th/0110146].
}
\lref\mooreseiberg{
  G.~W.~Moore and N.~Seiberg,
  ``Taming the Conformal Zoo,''
  Phys.\ Lett.\  B {\bf 220}, 422 (1989).
  }

\lref\csntworefs{
  B.~M.~Zupnik and D.~G.~Pak,
  ``Superfield formulation of the simplest three-dimensional gauge theories and
  conformal supergravities,''
  Theor.\ Math.\ Phys.\  {\bf 77}, 1070 (1988)
  [Teor.\ Mat.\ Fiz.\  {\bf 77}, 97 (1988)];
  E.~A.~Ivanov,
  ``Chern-Simons matter systems with manifest N=2 supersymmetry,''
  Phys.\ Lett.\  B {\bf 268}, 203 (1991);
  L.~V.~Avdeev, G.~V.~Grigorev and D.~I.~Kazakov,
  ``Renormalizations in Abelian Chern-Simons field theories with matter,''
  Nucl.\ Phys.\  B {\bf 382}, 561 (1992);
  L.~V.~Avdeev, D.~I.~Kazakov and I.~N.~Kondrashuk,
  ``Renormalizations in supersymmetric and nonsupersymmetric nonAbelian
  Chern-Simons field theories with matter,''
  Nucl.\ Phys.\  B {\bf 391}, 333 (1993).
}

\lref\nilssonpope{
  B.~E.~W.~Nilsson and C.~N.~Pope,
  ``Hopf Fibration Of Eleven-Dimensional Supergravity,''
  Class.\ Quant.\ Grav.\  {\bf 1}, 499 (1984).
}

\lref\SchnablWJ{
  M.~Schnabl and Y.~Tachikawa,
  ``Classification of N=6 superconformal theories of ABJM type,''
  arXiv:0807.1102 [hep-th].
}

\lref\WatamuraHJ{
  S.~Watamura,
  ``Spontaneous Compactification And $ CP^N: ~SU(3)\times  SU(2) \times U(1)$,
  $\sin^2\theta_W$, G(3) / G(2) And SU(3) Triplet Chiral Fermions In
  Four-Dimensions,''
  Phys.\ Lett.\  B {\bf 136}, 245 (1984).
}
\lref\HosomichiJD{
  K.~Hosomichi, K.~M.~Lee, S.~Lee, S.~Lee and J.~Park,
  ``N=4 Superconformal Chern-Simons Theories with Hyper and Twisted Hyper
  Multiplets,''
  arXiv:0805.3662 [hep-th].
}

\lref\GaiottoSD{
  D.~Gaiotto and E.~Witten,
  ``Janus Configurations, Chern-Simons Couplings, And The Theta-Angle in N=4
  Super Yang-Mills Theory,''
  arXiv:0804.2907 [hep-th].
}

\lref\gibbonshawking{
  G.~W.~Gibbons and S.~W.~Hawking,
  ``Gravitational Multi - Instantons,''
  Phys.\ Lett.\  B {\bf 78}, 430 (1978).
  }

\lref\hananybaryon{
  D.~Forcella, A.~Hanany, Y.~H.~He and A.~Zaffaroni,
  ``The Master Space of N=1 Gauge Theories,''
  arXiv:0801.1585 [hep-th].
}

\lref\hananyexp{
  B.~Feng, A.~Hanany and Y.~H.~He,
  ``Counting gauge invariants: The plethystic program,''
  JHEP {\bf 0703}, 090 (2007)
  [arXiv:hep-th/0701063].
}

\lref\indexref{
  J.~Kinney, J.~M.~Maldacena, S.~Minwalla and S.~Raju,
  ``An index for 4 dimensional super conformal theories,''
  Commun.\ Math.\ Phys.\  {\bf 275}, 209 (2007)
  [arXiv:hep-th/0510251].
}

\lref\MaldacenaSS{
  J.~M.~Maldacena, G.~W.~Moore and N.~Seiberg,
  ``D-brane charges in five-brane backgrounds,''
  JHEP {\bf 0110}, 005 (2001)
  [arXiv:hep-th/0108152].
}

\lref\AharonyQU{
  O.~Aharony and E.~Witten,
  ``Anti-de Sitter space and the center of the gauge group,''
  JHEP {\bf 9811}, 018 (1998)
  [arXiv:hep-th/9807205].
}

\lref\GubserFP{
  S.~S.~Gubser and I.~R.~Klebanov,
  ``Baryons and domain walls in an N = 1 superconformal gauge theory,''
  Phys.\ Rev.\  D {\bf 58}, 125025 (1998)
  [arXiv:hep-th/9808075].
}

\lref\BenaZB{
  I.~Bena,
  ``The M theory dual of a 3 dimensional theory with reduced supersymmetry,''
  Phys.\ Rev.\  D {\bf 62}, 126006 (2000)
  [arXiv:hep-th/0004142].
}


\lref\AharonyJU{
  O.~Aharony and A.~Hanany,
  ``Branes, superpotentials and superconformal fixed points,''
  Nucl.\ Phys.\  B {\bf 504}, 239 (1997)
  [arXiv:hep-th/9704170].
}

\lref\AlvarezGaumeIG{
  L.~Alvarez-Gaume and E.~Witten,
  ``Gravitational Anomalies,''
  Nucl.\ Phys.\  B {\bf 234}, 269 (1984).
}

\lref\SenZB{
  A.~Sen,
  ``Kaluza-Klein dyons in string theory,''
  Phys.\ Rev.\ Lett.\  {\bf 79}, 1619 (1997)
  [arXiv:hep-th/9705212].
}
\lref\RedlichDV{
  A.~N.~Redlich,
  ``Parity Violation And Gauge Noninvariance Of The Effective Gauge Field
  Action In Three-Dimensions,''
  Phys.\ Rev.\  D {\bf 29}, 2366 (1984).
}
\lref\KitaoMF{
  T.~Kitao, K.~Ohta and N.~Ohta,
  ``Three-dimensional gauge dynamics from brane configurations with
  (p,q)-fivebrane,''
  Nucl.\ Phys.\  B {\bf 539}, 79 (1999)
  [arXiv:hep-th/9808111].
}

\lref\HananyIE{
  A.~Hanany and E.~Witten,
  ``type IIB superstrings, BPS monopoles, and three-dimensional gauge
  dynamics,''
  Nucl.\ Phys.\  B {\bf 492}, 152 (1997)
  [arXiv:hep-th/9611230].
}

\lref\GW{
  D.~Gaiotto and E.~Witten,
  ``Janus Configurations, Chern-Simons Couplings, And The Theta-Angle in N=4
  Super Yang-Mills Theory,''
  arXiv:0804.2907 [hep-th].
}

\lref\multiplemtwo{
  J.~Gomis, G.~Milanesi and J.~G.~Russo,
  ``Bagger-Lambert Theory for General Lie Algebras,''
  arXiv:0805.1012 [hep-th].
  S.~Benvenuti, D.~Rodriguez-Gomez, E.~Tonni and H.~Verlinde,
  ``N=8 superconformal gauge theories and M2 branes,''
  arXiv:0805.1087 [hep-th].
  P.~M.~Ho, Y.~Imamura and Y.~Matsuo,
  ``M2 to D2 revisited,''
  arXiv:0805.1202 [hep-th].
  M.~A.~Bandres, A.~E.~Lipstein and J.~H.~Schwarz,
  ``Ghost-Free Superconformal Action for Multiple M2-Branes,''
  arXiv:0806.0054 [hep-th].
  J.~Gomis, D.~Rodriguez-Gomez, M.~Van Raamsdonk and H.~Verlinde,
  ``The Superconformal Gauge Theory on M2-Branes,''
  arXiv:0806.0738 [hep-th].
}

\lref\Polchinski{
J.~Polchinski, Volume 1.}


\lref\Shapere{
N.~J.~Evans, C.~V.~Johnson and A.~D.~Shapere,
  ``Orientifolds, branes, and duality of 4D gauge theories,''
  Nucl.\ Phys.\  B {\bf 505}, 251 (1997)
  [arXiv:hep-th/9703210].
}

\lref\WittenXY{
  E.~Witten,
  ``Baryons and branes in anti de Sitter space,''
  JHEP {\bf 9807}, 006 (1998)
  [arXiv:hep-th/9805112].
}

\lref\HananyKol{
  A.~Hanany and B.~Kol,
  ``On orientifolds, discrete torsion, branes and M theory,''
  JHEP {\bf 0006}, 013 (2000)
  [arXiv:hep-th/0003025].
}

\lref\HananyIE{
  A.~Hanany and E.~Witten,
  ``type IIB superstrings, BPS monopoles, and three-dimensional gauge
  dynamics,''
  Nucl.\ Phys.\  B {\bf 492}, 152 (1997)
  [arXiv:hep-th/9611230].
}

\lref\wittensl{
  E.~Witten,
  ``SL(2,Z) action on three-dimensional conformal field theories with Abelian
  symmetry,''
  arXiv:hep-th/0307041.
}

\lref\WittenDS{
  E.~Witten,
  ``Supersymmetric index of three-dimensional gauge theory,''
  arXiv:hep-th/9903005.
}

\lref\Sharpe{
  E.~R.~Sharpe,
  ``Analogues of discrete torsion for the M theory three-form,''
  Phys.\ Rev.\  D {\bf 68}, 126004 (2003)
  [arXiv:hep-th/0008170].
}

\lref\Sethi{
  J.~de Boer, R.~Dijkgraaf, K.~Hori, A.~Keurentjes, J.~Morgan, D.~R.~Morrison and S.~Sethi,
  ``Triples, fluxes, and strings,''
  Adv.\ Theor.\ Math.\ Phys.\  {\bf 4}, 995 (2002)
  [arXiv:hep-th/0103170].
}

\lref\Hosomichi{
  K.~Hosomichi, K.~M.~Lee, S.~Lee, S.~Lee and J.~Park,
  ``N=5,6 Superconformal Chern-Simons Theories and M2-branes on Orbifolds,''
  arXiv:0806.4977 [hep-th].
}

\lref\IntriligatorID{
  K.~A.~Intriligator and N.~Seiberg,
  ``Duality, monopoles, dyons, confinement and oblique confinement in
  supersymmetric SO(N(c)) gauge theories,''
  Nucl.\ Phys.\  B {\bf 444}, 125 (1995)
  [arXiv:hep-th/9503179].
}

\lref\GiveonSR{
  A.~Giveon and D.~Kutasov,
  ``Brane dynamics and gauge theory,''
  Rev.\ Mod.\ Phys.\  {\bf 71}, 983 (1999)
  [arXiv:hep-th/9802067].
}

\lref\Morrison{
  D.~R.~Morrison and M.~R.~Plesser,
  ``Non-spherical horizons. I,''
  Adv.\ Theor.\ Math.\ Phys.\  {\bf 3}, 1 (1999)
  [arXiv:hep-th/9810201].
}

\lref\WittenEY{
  E.~Witten,
  ``Dyons Of Charge E Theta/2 Pi,''
  Phys.\ Lett.\  B {\bf 86}, 283 (1979).
}

\lref\ABJM{
  O.~Aharony, O.~Bergman, D.~L.~Jafferis and J.~Maldacena,
  ``N=6 superconformal Chern-Simons-matter theories, M2-branes and their
  gravity duals,''
  arXiv:0806.1218 [hep-th].
}

\lref\BasuED{
  A.~Basu and J.~A.~Harvey,
  ``The M2-M5 brane system and a generalized Nahm's equation,''
  Nucl.\ Phys.\  B {\bf 713}, 136 (2005)
  [arXiv:hep-th/0412310].
}

\lref\GomisVC{
  J.~Gomis, D.~Rodriguez-Gomez, M.~Van Raamsdonk and H.~Verlinde,
  ``A Massive Study of M2-brane Proposals,''
  arXiv:0807.1074 [hep-th].
}



\Title{\vbox{\baselineskip12pt\hbox{}
\hbox{WIS/15/08-JUL-DPP}}} {\vbox{ \vskip -5cm
{\centerline{Fractional M2-branes}}}}

\vskip  -5mm
 \centerline{Ofer Aharony$^{a}$, Oren Bergman$^{b,c}$ and Daniel Louis Jafferis$^{d}$}

\bigskip
{\sl
\centerline{$^a$Department of Particle Physics}
\centerline{The Weizmann Institute of Science, Rehovot 76100, Israel}
\centerline{\tt Ofer.Aharony@weizmann.ac.il}
\medskip
\centerline{$^{b}$School of Natural Sciences, Institute for Advanced Study, Princeton, NJ 08540, USA}
\centerline{\tt bergman@sns.ias.edu}
\medskip
\centerline{$^{c}$Department of Physics, Technion, Haifa 32000, Israel}
\centerline{\tt bergman@physics.technion.ac.il}
\medskip
\centerline{$^{d}$Department of Physics, Rutgers University, Piscataway, NJ 08855, USA}
\centerline{\tt jafferis@physics.rutgers.edu}
\bigskip
\bigskip \medskip
}


\leftskip 8mm  \rightskip 8mm \baselineskip14pt \noindent
%
We consider two generalizations of the
${\cal N}=6$ superconformal Chern-Simons-matter theories with gauge
group $U(N)\times U(N)$. The first generalization is
to ${\cal N}=6$ superconformal $U(M)\times U(N)$ theories,
and the second to ${\cal N}=5$ superconformal
$O(2M)\times USp(2N)$ and $O(2M+1)\times USp(2N)$ theories.
These theories are conjectured to describe
M2-branes probing ${\bf C}^4/{\bf Z}_k$ in the unitary case,
and ${\bf C}^4/{\widehat{\bf D}}_k$ in the orthogonal/symplectic case,
together with a discrete flux, which can be interpreted as
$|M-N|$ fractional M2-branes localized at the orbifold singularity.
The classical theories with these gauge groups have been
constructed before; in this paper we focus on some quantum
aspects of these theories, and on a detailed description of their
M theory and type IIA string theory duals.

\bigskip\medskip

\leftskip 0mm  \rightskip 0mm
 \Date{\hskip 8mm July 2008}

\newsec{Introduction and summary of results}

Motivated by the work of Bagger and Lambert \baggerlambert\ (see also \refs{\schwarz\BasuED-\GustavssonVU}), there has recently been
great interest in studying three dimensional superconformal field theories and
their relation to the low-energy theory on M2-branes. The attempts to directly generalize
the Bagger-Lambert construction of an ${\cal N}=8$ superconformal theory with $SO(4)$ gauge group to the low-energy
theory on $N$ M2-branes have not yet been successful, and seem to lead either to non-unitary
theories or to non-conformal theories. In \ABJM\ we suggested an alternative route to
studying M2-branes. We presented a $U(N)_k \times U(N)_{-k}$ Chern-Simons matter theory with
explicit ${\cal N}=6$ superconformal symmetry\foot{These theories are special cases of ${\cal N}=4$ superconformal Chern-Simons theories constructed in \refs{\GaiottoSD,\HosomichiJD}.}. Considering the brane construction of this
theory led us to conjecture that it was equivalent to the low-energy theory of $N$ M2-branes
on a ${\bf C}^4/{\bf Z}_k$ singularity, and we presented various pieces of evidence for this
conjecture. In the special cases of $k=1$ and $k=2$ this conjecture implies that the theories
should have an enhanced ${\cal N}=8$ supersymmetry, whose generators cannot be described
locally in terms of the original fields of the theory. In particular, for $k=1$ we conjectured
that the $U(N)_1\times U(N)_{-1}$ theory is equivalent to the low-energy theory on $N$ M2-branes in flat space.

In this paper we generalize the construction of \ABJM\ in two directions, both of which
break the parity symmetry of the original construction. Both of these
generalizations have already been considered as classical
field theories in \Hosomichi, but we add more
details about quantum aspects of these theories, and we also discuss in detail their
gravitational duals. The first generalization we consider is to $U(M)_k\times U(N)_{-k}$
Chern-Simons-matter theories, with the same matter content and interactions as in \ABJM,
but with $M \neq N$. From the point of view of M2-branes probing a ${\bf C}^4/{\bf Z}_k$
singularity, these theories arise (for $M>N$) when we have $(M-N)$ fractional M2-branes
sitting at the singularity, together with $N$ M2-branes that are free to move around.
On the field theory side, classically this generalization is
straightforward, and still leads to ${\cal N}=6$ superconformal field theories.
However, we will argue (both directly in the field theory, and by using
a brane construction) that quantum mechanically these theories only exist as unitary
superconformal field theories when $|M-N| \leq |k|$. For these cases, we argue that the
gravitational dual is the same $AdS_4\times S^7/{\bf Z}_k$
background described in \ABJM, just with an additional
``torsion flux'', which takes values in $H^4(S^7/{\bf Z}_k,{\bf Z})={\bf Z}_k$.
This flux may be thought of as an M theory analog of orbifold discrete torsion,
corresponding to a discrete holonomy of the M theory 3-form field
on a torsion 3-cycle in $S^7/{\bf Z}_k$.
This is a good description of the gravitational dual when $N \gg k^5$.
For $k \ll N \ll k^5$, on the other hand, the appropriate description is in terms
of type IIA string theory on $AdS_4\times {\bf CP}^3$, with a discrete holonomy
of the NSNS 2-form field on the ${\bf CP}^1 \subset {\bf CP}^3$.
In this description the NSNS 2-form holonomy is quantized dynamically.

Our second generalization involves an orientifold in the original brane construction,
which changes the gauge symmetry to $O(M)\times USp(2N)$, and reduces the charged matter
content to two bi-fundamental chiral multiplets. The resulting gauge theories have ${\cal N}=5$
superconformal symmetry, and again there are bounds on $|M-2N|$ in order for the quantum
theory to exist (which we will describe in more detail below). We conjecture that these
theories are dual to the
low-energy
theory on
M2-branes probing a
${\bf C}^4/{\bf {\widehat D}}_k$ singularity, where
${\bf {\widehat D}}_k$ is the binary dihedral group with $4k$ elements, and
(generically) with some discrete flux, taking values in
$H^4(S^7/{\bf {\widehat D}}_k, {\bf Z}) = {\bf Z}_{4k}$, at the singularity.
As before, one can also think of this discrete flux as associated with fractional M2-branes
sitting at the singularity. The gravitational dual of these theories may
thus be described (for $N \gg k^5$) by M theory on $AdS_4\times S^7/{\bf {\widehat D}}_k$,
possibly with discrete torsion.
When $k \ll N \ll k^5$ this
background reduces to an orientifold of type IIA string theory on $AdS_4\times {\bf CP}^3$,
again with a discrete holonomy of the NSNS 2-form.

Our study provides additional examples of $AdS_4/CFT_3$ duals with extended supersymmetry,
which should hopefully be useful for understanding better the properties of this duality.
In particular, we extend the duality to theories which are not parity-invariant, and are
thus closer in spirit to the Chern-Simons-matter theories encountered in condensed matter
applications. It would be interesting to analyze these theories in the 't Hooft limit
(large $N$ and $k$ with fixed $N/k$) and
to see if integrability (on the field theory or string theory sides of the duality
\refs{\integrabilityft,\integrabilityst}) may be
used to learn about various properties of these theories.
Some of our results also have applications to more general three dimensional
theories. For instance, our claim (based on \bergman) that the $U(M)_k$ ${\cal N}=3$
supersymmetric pure Chern-Simons theories with $M>k$ do not
exist implies that almost all of the classical supersymmetric vacua of the
mass-deformed $U(N)\times U(N)$ theory found in \GomisVC\ do
not survive in the quantum theory (when $k=1$), so more vacua need to be found in order
to match with expectations\foot{We thank M. Van Raamsdonk for discussions on this issue.}.

\medskip

The rest of the paper is organized as follows.
In section 2 we consider the ${\cal N}=6$  $U(M)\times U(N)$ superconformal Chern-Simons-matter
theories, their realization in terms of brane configurations in type IIB string theory,
and their gravity dual descriptions.
In section 3 we consider the ${\cal N}=5$ $O(M)\times USp(2N)$ superconformal
theories, their type IIB brane realizations, and their gravity dual descriptions.
The moduli space of these theories is analyzed in an appendix.

\newsec{$U(M)\times U(N)$ theories}

It is straightforward to generalize the field theory construction of \ABJM\ to supersymmetric $U(M)_k \times
U(N)_{-k}$ Chern-Simons matter theories. As in the case of $N = M$, we can take
the ${\cal N}=3$ supersymmetric Chern-Simons-matter theory with this gauge group and with two
bi-fundamental hypermultiplets, and argue that because of the special form of
its superpotential, it in fact has an enhanced ${\cal N}=6$ supersymmetry
(these theories were also recently discussed in \Hosomichi).
Many aspects of the analysis
of these theories are very similar to the case of $N=M$ which was discussed in
\ABJM. For instance, the moduli space of these theories turns out to be exactly
the same as that of the $U(L)_k\times U(L)_{-k}$ theories
with $L = \min(N,M)$, and the
spectrum of non-baryonic chiral primary operators is also the same as the one in that theory. The main difference is that the effective field theory on
the moduli space contains an extra $U(M-N)_{k}$ Chern-Simons (CS) theory (when $M>N$), with
no massless charged fields.
In this section we will describe the gravitational dual of the $U(M)\times U(N)$
theories, and some of the differences between these theories and the $N=M$
theories.

\subsec{The field theories}

Let us assume to begin with that $M>N$, and consider the superconformal
theories $U(N+l)_k\times U(N)_{-k}$ with $l\geq 0$ (and $k > 0$).
Parity takes this theory to the $U(N)_k\times U(N+l)_{-k}$ theory.
We will see later that there is also an equivalence relating a theory of the first
type to a theory of the second type (with $l\rightarrow k-l$).
As in the equal rank case, the more general superconformal theory can be
obtained (at least classically)
as the IR limit of an ${\cal N}=3$ supersymmetric gauge theory, arising
as the low-energy limit of
a type IIB brane construction. Recall that the brane construction of the
$U(N)_k\times U(N)_{-k}$ theories involves \ABJM\ $N$ D3-branes winding around
a circle, and intersecting an NS5-brane and a $(1,k)$5-brane (a bound state of an NS5-brane with
$k$ D5-branes) at specific angles (described in \ABJM).
\ifig\fractionalbranes{The brane construction of the $U(N+l)_k\times U(N)_{-k}$ theory.}
{\epsfxsize2.5in\epsfbox{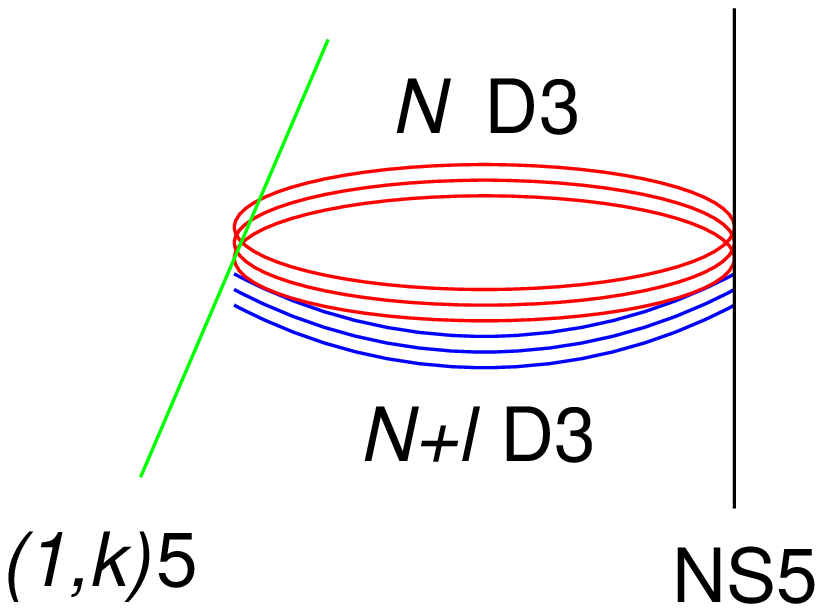}}
To get the $U(N+l)_k\times U(N)_{-k}$ theory we add to this
$l$ D3-branes which are suspended
between the NS5-brane and the $(1,k)$5-brane on one side
of the circle
(the side in which the
5-brane intersection numbers lead to a Chern-Simons level of $+k$,
see \fractionalbranes).
This construction makes it clear that the classical moduli space is identical to the moduli
space of the $U(N)_k\times U(N)_{-k}$ theory. It corresponds to the motion of the
$N$ free D3-branes (as well as to the Wilson line and dual gauge field on the
D3-branes). The $l$ suspended D3-branes are locked into position
by the two 5-branes, so there is no moduli space associated with them.

There is some additional ``braneology'' associated with these configurations,
which is reflected in interesting properties of the corresponding field theories
\foot{We thank A. Hashimoto for useful discussions on these brane configurations.}.
First, let us try to go onto the classical moduli space, by separating the $N$ branes
which wrap all the way around the circle from the other branes. This leaves only $l$
``fractional branes'' stretched between the two five-branes. For $l>k$, it was argued
in \bergman\ that this configuration of ``fractional branes'' breaks supersymmetry.
In the brane picture this follows from the ``s-rule'' \HananyIE\ forbidding
more than one D3-brane from ending on a specific NS5-brane D5-brane pair,
which was generalized to this case in \bergman.
From the field theory point of view,
at such a point on the moduli space
the low-energy field theory includes a pure ${\cal N}=3$ $U(l)_k$
YM-CS theory living on the fractional D3-branes.
A simple generalization of the Witten index computation
of \WittenDS\ implies that supersymmetry is unbroken for $l~\leq~k$.
However, for $l>k$, this computation suggests that supersymmetry is broken \bergman.
This implies that the classical moduli space of this $U(N+l)_k\times U(N)_{-k}$ brane
configuration is at least partly lifted when $l>k$.

\ifig\branecreation{Branes are created when we move a 5-brane around the circle.}
{\epsfxsize5in\epsfbox{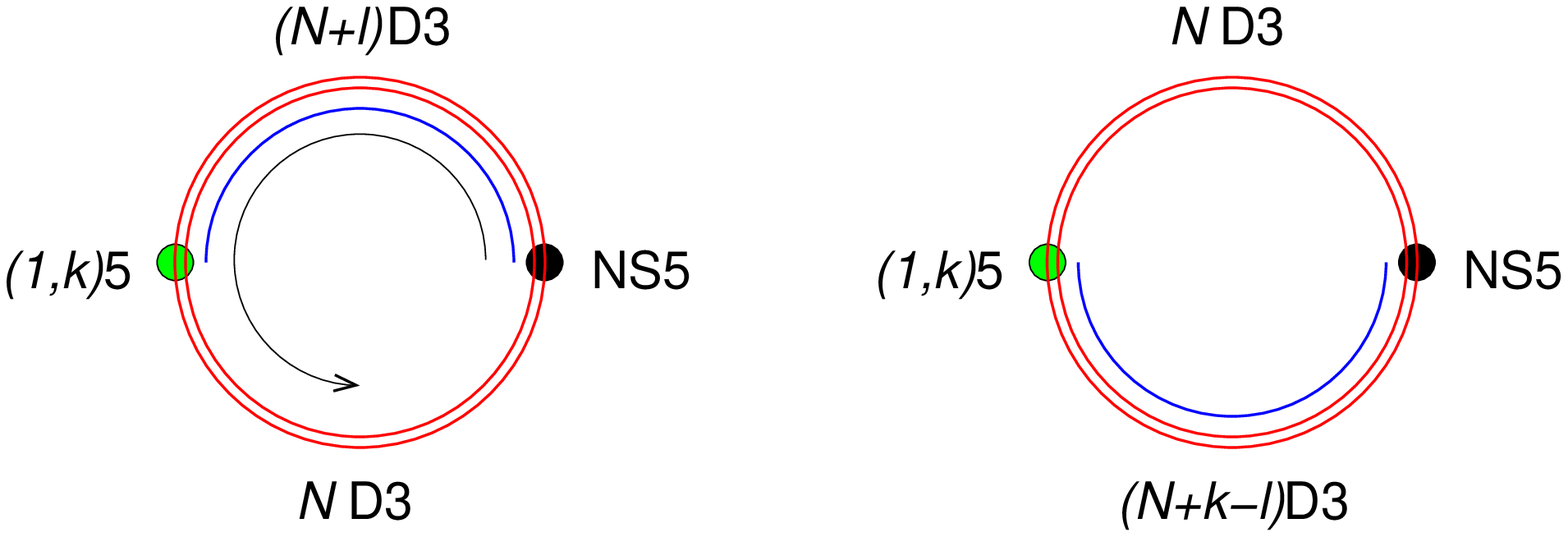}}

The second interesting bit of brane physics has to do with the
brane creation effect \HananyIE . When the NS5-brane crosses the $(1,k)$5-brane,
$k$ D3-branes are created between them, corresponding to one D3-brane
for each D5-brane component of the $(1,k)$5-brane.
If there are $l$ suspended D3-branes
to begin with, there will be $(k-l)$ suspended D3-branes after the
crossing. Let us start with the brane configuration for the ${\cal N}=3$ gauge
theory with $U(N+l)_k\times U(N)_{-k}$ where $0\leq l\leq k$, and move
one of the 5-branes around the circle
in the direction that initially shortens the suspended 3-branes.
The final configuration then describes the ${\cal N}=3$ gauge theory with
$U(N)_k\times U(N+k-l)_{-k}$, see \branecreation.
We thus expect these two gauge theories to flow to the same superconformal theory.
Below we will argue that these two superconformal theories are indeed equivalent.
\foot{Naively, moving the 5-brane around the circle multiple times, or in the other direction,
should produce more equivalences. However, all of these additional configurations lead
either to anti-D3-branes or to $U(N'+l')_k \times U(N')_{-k}$ theories with $l' > k$.}

Let us now consider the ${\cal N}=6$ superconformal
$U(N+l)_k\times U(N)_{-k}$ Chern-Simons-matter theory that is obtained
in the IR limit of the above gauge theory. We will use the description of these theories
in \ABJM, in which they are written as special cases of ${\cal N}=3$ supersymmetric
Chern-Simons-matter theories that happen to have enhanced supersymmetry.
We will now argue that the theories with $l>k$ do not exist as unitary conformal
field theories, and that the above theory with $l\leq k$ is equivalent
to the theory with $U(N)_k\times U(N+k-l)_{-k}$.

In Chern-Simons-matter theories with this much supersymmetry the metric on the moduli
space cannot be corrected, so we can go
to a generic point on the moduli space, where at low energy we get
$N$ copies of a $U(1)_k\times U(1)_{-k}$ theory (as in the $U(N)_k\times U(N)_{-k}$ case),
together with a pure $U(l)_k$ ${\cal N}=3$ supersymmetric Chern-Simons theory, with no massless charged matter fields.
Naively, this supersymmetric Chern-Simons theory is the same as the bosonic Chern-Simons theory at level $k$, since the other fields in the ${\cal N}=3$ vector multiplet just have Gaussian
actions in the absence of any matter fields, but in fact we should be more careful.
Recall that the Chern-Simons level in \ABJM\ was defined such that it
was equal to the level of the ${\cal N}=3$ Yang-Mills-Chern-Simons theory
which flows to the Chern-Simons-matter theory in the IR. In this theory the level is not
renormalized \cscoefficient, but this comes from the cancelation of two contributions.
There is a shift of $k \to k - l \cdot {\rm sign}(k)$ coming from integrating out the gauginos,
and another opposite shift from the
one-loop contributions of the gauge field itself (these shifts are only in the $SU(l)$ level,
not in the $U(1)$ level, but this will not affect our arguments). Thus, if we think of this theory in bosonic terms, we can say that it includes a bosonic $SU(l)_{k-l\cdot {\rm sign}(k)}$ Chern-Simons theory.
However this only makes sense if $|k| \geq l$, since otherwise the one-loop shift
in the bosonic Chern-Simons theory (which generally takes $k \to k + l \cdot {\rm sign}(k)$) would have the wrong sign and would not bring us back to
the level $k$ theory that we want. This suggests that the ${\cal N}=3$ supersymmetric
$U(l)_k$
pure Chern-Simons theory does not exist as a unitary theory for $l>k$ (at least, it does not
seem to be equivalent to any standard bosonic Chern-Simons theory).
We are therefore led to conjecture that the superconformal $U(N+l)_k\times U(N)_{-k}$
theories with $l>k$ do not exist as unitary theories. Note that these theories are never
weakly coupled (recall that a $U(N)_k$ theory is weakly coupled only when $|k| \gg N$).
This conjecture is consistent with what we found in the corresponding ${\cal N}=3$ $U(N+l)_k\times U(N)_{-k}$
Yang-Mills-Chern-Simons theories, where we showed that for $l>k$ the moduli space was
partly lifted, so these theories cannot flow to the $U(N+l)\times U(N)$ Chern-Simons-matter
theories (whose moduli space does not receive quantum corrections).

For the superconformal theories with $l\leq k$ the brane picture suggested that
there
should be
an equivalence between pairs of theories of the form
\eqn\equivalence{
U(N+l)_{k}\times U(N)_{-k} = U(N)_k \times U(N+k-l)_{-k} .}
Note that again at least one of these two theories is always strongly coupled,
so it is not easy to verify this duality\foot{In particular, these
theories seem to have a
different number of degrees of freedom, but this number
can be significantly modified at strong coupling.}.
As (weak) evidence for this conjectured equivalence, note that on the
moduli space one obtains the $U(l)_k$ supersymmetric CS theory
for the theory on the left-hand side of \equivalence, and
the $U(k-l)_{-k}$ theory for the theory on the right-hand side.
The corresponding effective bosonic theories therefore include
pure $SU(l)_{k-l}$ CS and pure $SU(k-l)_{-l}$ CS, respectively.
In the presence of a boundary, we obtain the corresponding WZW theories,
$SU(l)_{k-l}$ and $SU(k-l)_l$,
which are precisely related by level-rank duality.
For the special case $l=k$ we get an equivalence between the
$U(N+k)_k\times U(N)_{-k}$ theory and the original $l=0$ theory
with $U(N)_k\times U(N)_{-k}$.
Note that in this case on the moduli space of the former theory
we seem to obtain an extra $U(k)_k$ pure Chern-Simons theory, but as argued above, at
least the $SU(k)$ part of this is equivalent to a bosonic $SU(k)_0$ theory, which is an
empty theory\foot{There is also a remaining $U(1)_k$ pure CS theory.
We will see below that in their dual description these two theories differ
by a shift of a bulk theta angle by $2\pi$. The extra $U(1)_k$ part may be related to the
fact that the boundary
theory changes by a $U(1)$ Chern-Simons theory
when the theta angle in the bulk is shifted by $2\pi$ \wittensl.}.
We therefore end up with a total of $k$
inequivalent theories for any given minimal rank $N$ (which sets the dimension of the
moduli space to be $8N$).
Below we will see that this is consistent with the dual
gravitational picture.
Note that, as expected, for the case $k=1$ corresponding to M2-branes in flat space
we do not find any new theories associated with ``fractional branes''.

\subsec{M theory lift}

Lifting the type IIB brane configuration for the $U(N+l)_k\times U(N)_{-k}$ theory
to M theory as in \ABJM, we obtain $N$ M2-branes moving in ${\bf C}^4/{\bf Z}_k$,
together with $l$ ``fractional M2-branes'', which correspond to M5-branes
wrapped on a vanishing 3-cycle at the orbifold point.
This is a pure torsion cycle since $H_3(S^7/{\bf Z}_k,{\bf Z})={\bf Z}_k$.
In other words, $k$ wrapped M5-branes are equivalent to none,
and there are $k$ inequivalent  configurations.
This can be thought of as an M theory analog of
orbifold discrete torsion \Sharpe ,
classified by $H^3(\Gamma,U(1))$, where $\Gamma = {\bf Z}_k$. This discrete
torsion may be
realized by a discrete holonomy of the 3-form potential on the above
3-cycle, $\exp(i\int_{3-cycle}C_3) \in {\bf Z}_k$.
By Poincar\'e duality we can also associate the fractional branes
to a pure torsion flux of the 4-form field strength in
$H^4(S^7/{\bf Z}_k,{\bf Z})={\bf Z}_k$.
We therefore conclude that the $U(N+l)_k\times U(N)_{-k}$ superconformal
theory with $0\leq l \leq k-1$ describes $N$ M2-branes on ${\bf C}^4/{\bf Z}_k$
with $l$ units of discrete torsion.
Our analysis of the discrete torsion
generalizes the known result in the case of $k=2$, where the two
variants of ${\bf C}^4/{\bf Z}_2$ are the $OM2^+$ and $OM2^-$ orbifold planes
\refs{\SethiZK,\HananyKol}.

Note that the theory of two M2-branes on the
${\bf C}^4/{\bf Z}_2$ orbifold with discrete torsion was also claimed \refs{\dmpv,\lamberttong}
to be described by the ${\cal N}=8$ Bagger-Lambert theory, with gauge group
$SU(2)\times SU(2)$, at level $k=2$. Here we claim that the same theory may
be described using the ${\cal N}=6$ $U(3)_2 \times U(2)_{-2}$ Chern-Simons-matter theory, which
has a hidden ${\cal N}=8$ superconformal symmetry (as well as a hidden parity
symmetry).

The number of discrete torsion variants is consistent with the restriction
$l\leq k$, together with the equivalence of the $l=0$ and $l=k$ theories,
found above for the superconformal field theories.
In fact the more general equivalence between the pairs of theories in
\equivalence\ can be understood as a parity symmetry in M theory,
that reflects one of the coordinates along the M2-branes and at the same time
takes $C_3 \rightarrow - C_3$.
The former has the effect of changing the sign of the Chern-Simons
term, and therefore the level $k\rightarrow -k$,
and the latter has the effect of changing the torsion
class $[G_4]\rightarrow k-[G_4]$, and therefore $l\rightarrow k-l$
in the gauge group that has level $(-k)$.

\subsec{M theory dual}

Taking the near-horizon limit of the brane configuration described above as in \ABJM,
we find that our field theories are dual to M theory on
$AdS_4\times S^7 / {\bf Z}_k$, with $N$ units of 4-form flux in $AdS_4$,
and $l$ units of discrete torsion:
\eqn\elevnd{ \eqalign{
ds^2 =&  \, { R^2 \over 4 } ds^2_{AdS_4} + R^2 ds^2_{S^7/{\bf Z}_k},
\cr
G_4 \sim  & \, N \epsilon_4,
\cr
\int_{S^3/{\bf Z}_k \subset S^7/{\bf Z}_k} C_3 = & \, 2\pi {l\over k}\,,
}}
where $\epsilon_4$ is the unit volume form on $AdS_4$
and $R = (2^5 \pi^2 k N)^{1/6} l_p$.
For $N \gg k^5$ the
theory has a good approximation in terms of eleven dimensional supergravity on
this background. We can describe $S^7/{\bf Z}_k$ in terms of 4 complex coordinates
$z_i$ with a constraint $\sum |z_i|^2=1$ and an identification $z_i \sim
e^{2\pi i / k} z_i$. In this language,
the 3-cycle $S^3/{\bf Z}_k$ can be described (for instance)
as $\{|z_1|^2+|z_2|^2=1, z_3=z_4=0\}$.

We can also describe the $S^7/{\bf Z}_k$ as an $S^1$ fibred over
${\bf CP}^3$, in which case its metric is expressed as
 \eqn\nermetr{
 ds^2_{S^7/{\bf Z}_k} = { 1 \over k^2 } ( d \varphi + k \omega)^2 + ds^2_{CP^3},
 }
where $\varphi\sim\varphi + 2\pi$, and $d\omega = J$ where $J$ is the K\"ahler
form on ${\bf CP}^3$.
The 3-form which has the above holonomy can then be expressed locally
as
\eqn\localthreeform{
C_3 \propto {l\over k} J\wedge d\varphi \,.
}

\subsec{Type IIA string theory dual}

Recall that for $M=N$, when $N^{1/5}\ll k \ll N$, the M theory circle becomes small (in eleven dimensional
Planck units). A better
description of the background dual to the $U(N)_k\times U(N)_{-k}$ theory is obtained \ABJM\ by reducing to type IIA string theory on
$AdS_4\times {\bf CP}^3$,
with $N$ units of $\tilde{F}_4$ flux on $AdS_4$ and
$k$ units of $F_2$ flux on the ${\bf CP}^1$ 2-cycle in ${\bf CP}^3$;
we can write this
background (setting the string scale to one) as
 \eqn\twoasol{ \eqalign{
 ds^2_{string} = & { R^3 \over k} ( { 1 \over 4 } ds^2_{AdS_4} + ds^2_{CP^3 } ),
 \cr
 e^{2 \phi} = & { R^3 \over k^3 } \sim { N^{1/2} \over k^{5/2} }= { 1 \over N^2 } \left( {
 N \over k } \right)^{5/2},
 \cr
 \tilde{F}_{4} = & { 3 \over 8 }  {  R^3} \hat \epsilon_4 ,
 \cr
 F_2 = & k d \omega = k J ,
 }}
where the radius of curvature in string units is
$R^2_{str}=R^3/k=2^{5/2} \pi \sqrt{N/k}$.

To describe the $U(N+l)_k\times U(N)_{-k}$ theories we need to add to this background
the reduction of the discrete torsion in M theory.
The 3-form in \localthreeform\ has one leg on the M theory circle, so it reduces
to the NSNS 2-form $B_2$, which attains a non-trivial holonomy on the
${\bf CP}^1$ 2-cycle in ${\bf CP}^3$
\eqn\twoformflux{
b_2 \equiv {1\over {2\pi}} \int_{{\bf CP}^1 \subset {\bf CP}^3} B_2 = {l\over k}\,.
}
Note that we are using here units in which $b_2$ is periodic with period one.
Equation \twoformflux\ seems like a surprising result at first, since the holonomy of $B_2$ is
a continuous variable, {\it i.e.} there is no discrete torsion in ${\bf CP}^3$.
Let us therefore explain how it arises directly in the type IIA description
of the fractional branes.

The M5-branes wrapped on $S^3/{\bf Z}_k$ reduce to D4-branes wrapped on
the ${\bf CP}^1$ 2-cycle of ${\bf CP}^3$.
Note that the low-energy theory on $l$ such D4-branes
contains a $U(l)$ Chern-Simons term
at level $k$, as we expect, due to the $F_2\wedge A \wedge dA$ term in
the D4-brane effective action (where $A$ is the D4-brane worldvolume gauge field).
From the point of view of the theory on $AdS_4$, the D4-branes are domain walls
separating the space into two separate regions. Of course, these domain walls (which
we think of as filling space and sitting at a fixed radial position) are not static, but
rather they feel a force driving them towards the horizon of $AdS_4$; this corresponds to
the fact that there is no moduli space for moving the fractional branes around. Nevertheless,
let us assume for a moment that we fix the branes at some fixed radial position. The D4-branes
are a source for the RR 5-form field $C_5$; this source means that as we go from one side of
the D4-branes to the other, the flux of the RR 4-form field strength $\tilde{F}_4$,
integrated over ${\bf CP}^2 \subset {\bf CP}^3$ (which is the dual cycle to the one the branes
are wrapped on), jumps by $l$ units.

Naively this suggests that the type IIA dual of the $U(N+l)\times U(N)$ theories should be described
by a configuration with $l$ units of 4-form flux on ${\bf CP}^2 \subset {\bf CP}^3$.
However, such a solution does not seem
to exist, and we wish to claim that the correct background is actually the one described above.
The point is that we also have $k$ units of $F_2$ flux on ${\bf CP}^1$, and in the presence
of this flux the equation of motion of ${\tilde F}_4$ is modified to
$d {\tilde F}_4 = - F_2 \wedge H_3$.\foot{Recall that the gauge-invariant 4-form field strength in type IIA
string theory is given by $\tilde{F}_4 = dC_3 - C_1\wedge H_3$.}
From the point of view of the effective field theory
on $AdS_4$, this means that $f_4 \propto \int_{{\bf CP}^2} {\tilde F}_4$
(normalized so that it is quantized to be an integer)
is not conserved, but
rather its equation of motion is given by $d(f_4 + k b_2) = 0$.
This means that the conserved flux which jumps by $l$ units as we cross the
D4-branes is not $f_4$ but rather $f_4 + k b_2$, and another way to realize such a jump is
by having $b_2$ jump by $l/k$.
This is precisely what we obtained above. Thus, we conjecture that
the type IIA string theory dual of the $U(N+l)_k \times U(N)_{-k}$ theory is
the original $AdS_4\times {\bf CP}^3$ background, together with
a $B$ field holonomy $b_2 = l/k$.
Clearly this is a solution of the classical equations of motion of type IIA supergravity
(preserving the same supersymmetries as the original background),
since $b_2$ does not enter the equations.

Note that naively $b_2$ is not quantized, but the arguments above suggest that
$(f_4 + k b_2)$
is quantized, implying that when $f_4=0$, $b_2$ must be quantized in units of $1/k$.
The periodicity $b_2\sim b_2 + 1$
then implies that there are $k$ inequivalent theories,
in agreement with both M theory and the field theory arguments. As a consistency check,
note that the equivalence of $b_2 \to b_2+1$ naively comes from the presence of
NS5-branes wrapped on ${\bf CP}^2 \subset {\bf CP}^3$, which from the $AdS_4$ point of
view look like axionic strings such that as we go around them, $b_2 \to b_2+1$. However,
in the presence of $k$ units of $F_2$
flux, such a wrapped NS5-brane must have $k$ D4-branes wrapped on
${\bf CP}^1$ ending on it. This is consistent with the fact that inserting $k$ fractional D4-branes
is the same as taking $b_2 \to b_2+1$, since such D4-branes can annihilate to nothing
by creating bubbles of wrapped NS5-branes which they can end on, and these bubbles can then
shrink to nothing, leaving behind only the
change $b_2 \to b_2+1$.

Furthermore, the type IIA version of the parity symmetry is given by the reflection of a coordinate
in $AdS_4$ together with $b_2\rightarrow -b_2$. Combining this with $b_2\rightarrow b_2 +1$
we find that the theory with $b_2 = l / k$ should be related to the theory with
$b_2 = (k - l) / k$ by a parity transformation, implying that the $U(N+l)_k \times U(N)_{-k}$
theory should be equivalent to the $U(N)_k\times U(N+k-l)_{-k}$ theory.
This is also in complete agreement with the field theory and M theory points of view
described above.

\subsec{A simple test}

Our derivation of the M theory / string theory duals was of course far from rigorous, so we should now make various
tests to see if our conjecture is sensible or not. First, note that the moduli space for a
single M2-brane in our M theory background is still given by ${\bf C}^4 / {\bf Z}_k$, in agreement
with the moduli space we expect. Similarly, the spectrum of light fields in our
backgrounds is not affected by the holonomy of the $C$ field or of the $B$ field, in agreement with our claim that the spectrum
of (non-baryonic) chiral primary operators in these theories is independent of $l$.

What effect does this holonomy have at all, given that it does not modify the classical
equations of motion? It does change the spectrum of baryonic
operators. Let us analyze this explicitly using the type IIA description.
Recall that the background above
allows for D4-branes wrapped on ${\bf CP}^2 \subset {\bf CP}^3$, which were identified in
\ABJM\ with di-baryon operators\foot{Note that these operators are not gauge-invariant
 under one of the $U(1)$ gauge groups; nevertheless they are present in the bulk. Presumably
 one way to think of these operators is like electrons in QED,
 as defined together with a Wilson
 line that carries away the $U(1)$ charge to infinity.} of the form $(C_I)^N$, where $C_I$ are bi-fundamentals
of $U(N)\times U(N)$. The worldvolume action of the D4-branes contains a coupling
$B_2\wedge F_2\wedge A$, implying that when $b_2 = l/k$, we must have $l$ strings
ending on the D4-brane. This is exactly what we expect for the di-baryon operators in the
$U(N+l)\times U(N)$ theory, since if we take an operator like $(C_I)^N$, we can contract
the $U(N)$ indices to a singlet of $SU(N)$ with an epsilon symbol, but if we then contract
the indices of $SU(N+l)$ by an epsilon symbol we are left with an object in the $l$'th
anti-symmetric product of fundamentals of $SU(N+l)$. This exactly matches the number of
strings that must end on this object to form a consistent state. (Of course, the operator
$(C_I)^{N+l}$ similarly transforms in the $l$'th anti-symmetric product of fundamentals
of $SU(N)$.) Thus, the spectrum of di-baryons exactly matches what we expect.

However, we have been a bit too fast in the analysis above; what really appears in the
source term on the D4-branes used above is $F_2 \wedge ({\hat F}_2-B_2)\wedge A$, where
${\hat F}_2$ is the gauge field strength on the D4-branes, and usually we say that in
the presence of a $B$ field, ${\hat F}_2-B_2$ is quantized rather than ${\hat F}_2$, so
one might suspect that we must turn on some ${\hat F}_2$ gauge field to cancel the
above effect of the $B$ field. We claim that in our background this is not correct, and
we are still allowed to have configurations with ${\hat F}_2=0$ that lead to the di-baryons
described above. To see this, let us analyze the effect of the $B$ field in a different
way, by looking at the four dimensional effective action on $AdS_4$. As we discussed, the
$B$ field does not affect the equations of motion, but it does affect the action. The ten
dimensional action includes a term $B_2 \wedge {\tilde F}_4 \wedge {\tilde F}_4$. So,
when we reduce to four dimensions, and look at the gauge field coming from $\int_{{\bf CP}^1} {\tilde F}_4$, the $b_2$ component of the $B$
field behaves as a theta angle for this gauge field. We know that when
we shift the theta angle, the Witten effect \WittenEY\ implies that the electric charges of particles
are shifted by the theta angle times their magnetic charges. In the case of the specific
gauge field we are discussing, the electrically charged particles are D2-branes wrapped on
${\bf CP}^1$, while the magnetically charged particles are D4-branes wrapped on ${\bf CP}^2$.
So, we expect that turning on the $B$ field will change the wrapped D2-brane charge of a wrapped D4-brane by an amount proportional to the $B$ field. And indeed, this is exactly what happens,
because of the $B_2 \wedge C_3$ coupling on the worldvolume of the D4-branes. Note that again
this coupling is really of the form $({\hat F}_2 - B_2)\wedge C_3$, so that we would not get
this effect if $({\hat F}_2-B_2)$ were quantized; but the space-time argument above implies
that we should see this effect, so we conclude that D4-brane configurations with ${\hat F}_2=0$ are
still allowed\foot{This is not necessarily true for other wrapped D-branes.}, and reproduce for us the expected di-baryon spectrum discussed above. It would be interesting to reproduce this
result by a direct study of the quantization of ${\hat F}_2$ in our background.

\subsec{Additional comments}

Our conjectured duality implies that at strong coupling (large $\lambda = N/k$),
all the $U(M)_k\times U(N)_{-k}$ theories with $M=N, N+1, \cdots, N+k$ are very similar to
each other, since they only differ by the $B$ field, which only affects the D-brane spectrum
and worldsheet instanton effects (which are suppressed by $\exp(-\sqrt{\lambda})$). In
particular, it seems that all these theories should be identical (at large $N$) to all orders in
perturbation theory in $1/\sqrt{\lambda}$, though not beyond this perturbation theory. For
instance, the entropy of these theories should be the same to all orders in $1/\sqrt{\lambda}$. Of
course, it is hard to check this claim, since we do not know how to perform computations
in these theories at strong coupling.

Many of the computations that were performed for the $U(N)_k\times U(N)_{-k}$
theories can easily be generalized to the $U(M)_k\times U(N)_{-k}$ theories. In particular,
it would be interesting to examine the spin chain representation of the spectrum of operators
in these theories in the 't Hooft limit \integrabilityft,
and to see if $(M-N)$ can be related to a $B$ field
on the ``worldsheet'' of the spin chain.

\newsec{Orientifold theories}

A class of ${\cal N}=5$ superconformal Chern-Simons-matter
theories with gauge groups $O(M)\times USp(2N)$ was recently
constructed in \Hosomichi. We will review how one can obtain these
theories as orientifold projections of the ${\cal N}=6$
superconformal theories, or alternatively as the IR limits of
gauge theories living on D3-branes in a type IIB orientifold-brane
configuration\foot{A different orientifold of these theories, which
breaks supersymmetry, and the corresponding brane construction, were
recently discussed in \ArmoniKR.}. We will argue that these theories describe
M2-branes probing the orbifold ${\bf C}^4/\widehat{\bf D}_k$,
where $\widehat{\bf D}_k$ is the binary dihedral group with $4k$
elements, together
with discrete torsion (or ``fractional M2-branes"). The dual
description (which is useful for $N,M \gg k$) is then either M
theory on $AdS_4\times S^7/\widehat{\bf D}_k$, or type IIA string
theory on $AdS_4\times {\bf CP}^3/{\bf Z}_2$, where the ${\bf
Z}_2$ is an orientifold projection that we will describe below.

\subsec{Type IIB orientifold brane configurations}

Let us begin with the type IIB brane configurations.
The orientifold theories are obtained by including an orientifold 3-plane
wrapped on the circle, in addition to the D3-branes and the two 5-branes
that we had in the previous section. Adding the orientifolds does not break
any additional supersymmetry.
We will take the D3-branes to be ``whole 3-branes" in the sense that they have
images in the covering space of the orientifold
(they are not identified with their own image).
The 5-branes on the other hand intersect the orientifold plane,
and are their own images. They are therefore ``half branes" in a sense.

The orientifold 3-plane comes in four varieties, which are denoted
O$3^-$, O$3^+$, $\widetilde{{\rm O} 3}^-$ and $\widetilde{{\rm O} 3}^+$.
When we have $N$ D3-branes sitting on top of the orientifold 3-plane,
these lead to the gauge groups
$O(2N)$, $USp(2N)$, $O(2N+1)$ and $USp(2N)$, respectively.
The O3-planes carry fractional D3-brane charges given by $(-1/4)$
for O$3^-$ and $(+1/4)$ for O$3^+$, $\widetilde{{\rm O} 3}^-$
and $\widetilde{{\rm O} 3}^+$.
In particular, the $\widetilde{{\rm O} 3}^-$ plane can be thought of
as an O$3^-$ plane with a half D3-brane stuck on top of it.
The two $USp(2N)$ theories corresponding to O$3^+$ and $\widetilde{{\rm O} 3}^+$
are perturbatively identical, but differ in their non-perturbative (dyon) spectrum.
The four types of orientifolds are related by the $SL(2,{\bf Z})$ duality of type IIB
string theory \WittenXY .
The O3-plane variants also correspond to a choice of discrete torsion
for the NSNS and RR 3-form field strengths, which take
values in $H^3({\bf RP}^5,\widetilde{\bf Z})={\bf Z}_2$ \WittenXY .
In particular, NSNS torsion differentiates the $(-)$ and $(+)$ variants,
and RR torsion differentiates the tilde and no-tilde variants.
A single 5-brane that intersects an orientifold 3-plane in two spatial
dimensions will therefore change it from one type on one side to
a different type on the other side \refs{\Shapere , \HananyKol} .
In particular, crossing an NS5-brane changes O$3^-$ to O$3^+$
and $\widetilde{{\rm O} 3}^-$ to $\widetilde{{\rm O} 3}^+$,
and crossing a D5-brane changes O$3^-$ to $\widetilde{{\rm O} 3}^-$
and O$3^+$ to $\widetilde{{\rm O} 3}^+$.

\ifig\fourclassesfig{The brane constructions of the orientifold theories.}
{\epsfxsize4in\epsfbox{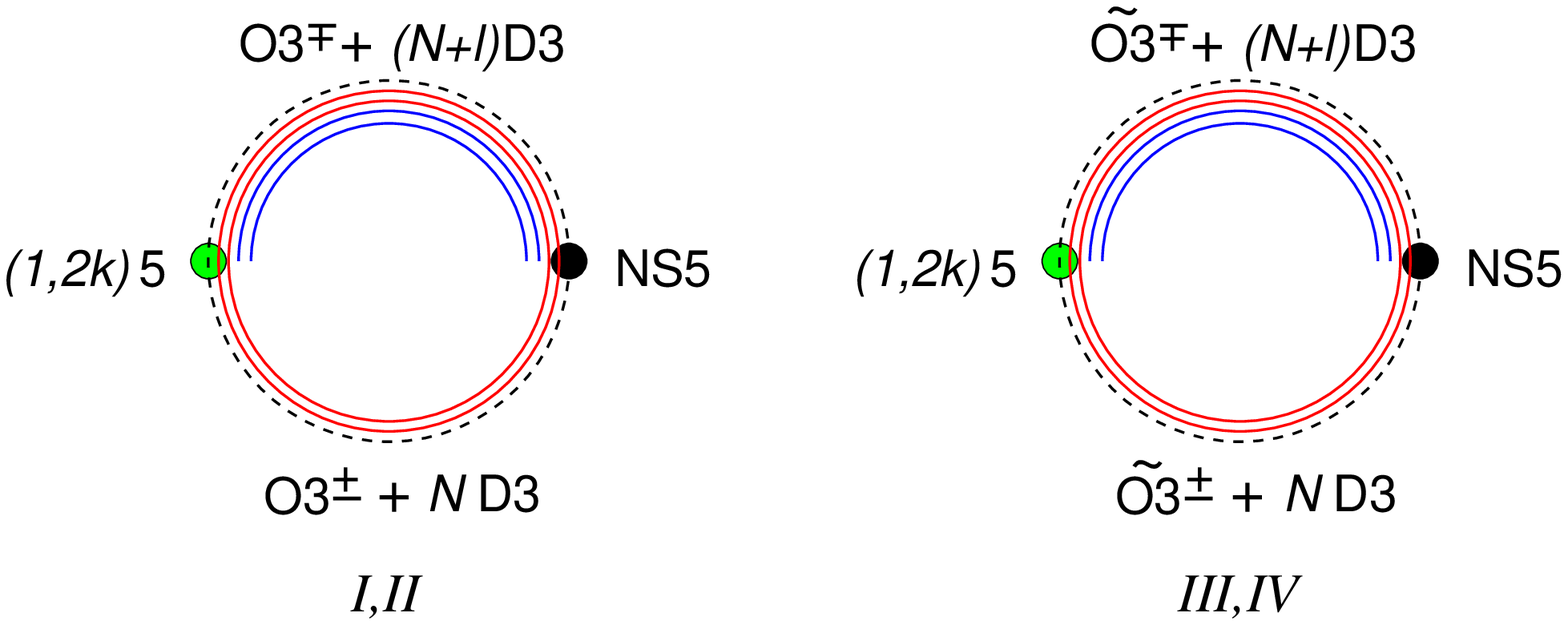}}

In our circle configuration we therefore get an orthogonal gauge group on one interval
between the two 5-branes, and a symplectic gauge group on the other interval.
For consistency of the configuration the D5-brane charge of the second 5-brane must be even,
so we will take it to be $(1,2k)$. Otherwise, the discrete torsion in the brane
configuration is not single-valued.
The $O(M)$ gauge field then has a Chern-Simons term at level $\pm 2k$,
and the $USp(2N)$ gauge field has a Chern-Simons term at level
$\mp k$.\foot{This gives another explanation of why we must start with $2k$ units
of D5-brane charge. The level of the $USp(2N)$ Chern-Simons term must
be an integer.}
We therefore find four classes of theories (see \fourclassesfig )
\eqn\fourclasses{
\eqalign{
I  & \quad O(2N+2l_I)_{2k}\times USp(2N)_{-k}, \cr
II & \quad USp(2N+2l_{II})_k\times O(2N)_{-2k}, \cr
III & \quad O(2N+2l_{III}+1)_{2k}\times USp(2N)_{-k}, \cr
IV & \quad USp(2N+2l_{IV})_k\times O(2N+1)_{-2k},
}}
and four more which are obtained by a parity transformation that takes
$k\rightarrow -k$.
We will use a shorthand notation to denote each of these theories,
for example the theory in class $I$ at level $k$ with $l$ fractional branes
will be denoted as $I_k(l)$.
The matter content can be determined by the action of the orientifold projection
on the original $U(2(N+l))\times U(2N)$ bi-fundamental fields $C_I$ ($I=1,2,3,4$) :
\eqn\orientifoldmatter{
C_{Ia\bar{b}} \rightarrow - M_{IJ} C_{Ja\bar{c}}^* J_{\bar{c}\bar{b}} \, ,
}
where $J$ is the invariant anti-symmetric matrix of the $USp$ theory,
and $M_{IJ}$ is a matrix acting on the four complex scalars (and their superpartners)
as $i\sigma_2 \otimes {\bf 1}_{2\times 2}$.
This projects out half of the two bi-fundamental hypermultiplets that we started from,
leaving a single bi-fundamental hypermultiplet, or equivalently
two bi-fundamental chiral superfields.
For example, choosing a specific ordering of the $C_I$, we can
take the identification on the chiral superfields $A_i$ and $B_i$
described in \ABJM\ to take the form $A_1 \equiv B_1^T J$, $A_2
\equiv B_2^T J$. We therefore obtain an ${\cal N}=3$ YM-CS theory
with two chiral bi-fundamental superfields $A_1$ and $A_2$, and
with a product of an orthogonal and symplectic gauge group in one
of the classes shown in \fourclasses , or their parity partners.
The superpotential is the projection of the original
superpotential $W \propto \tr(A_1 B_1 A_2 B_2 - A_1 B_2 A_2 B_1)$
using the identifications above, which gives \eqn\orientsuperpot{W
\propto \tr(A_1 J A_1^T A_2 J A_2^T - A_1 J A_2^T A_2 J A_1^T).}

\ifig\newbranecreation{A D3-brane is created when two NS5-branes cross on an
O$3^\mp$-plane.}
{\epsfxsize4in\epsfbox{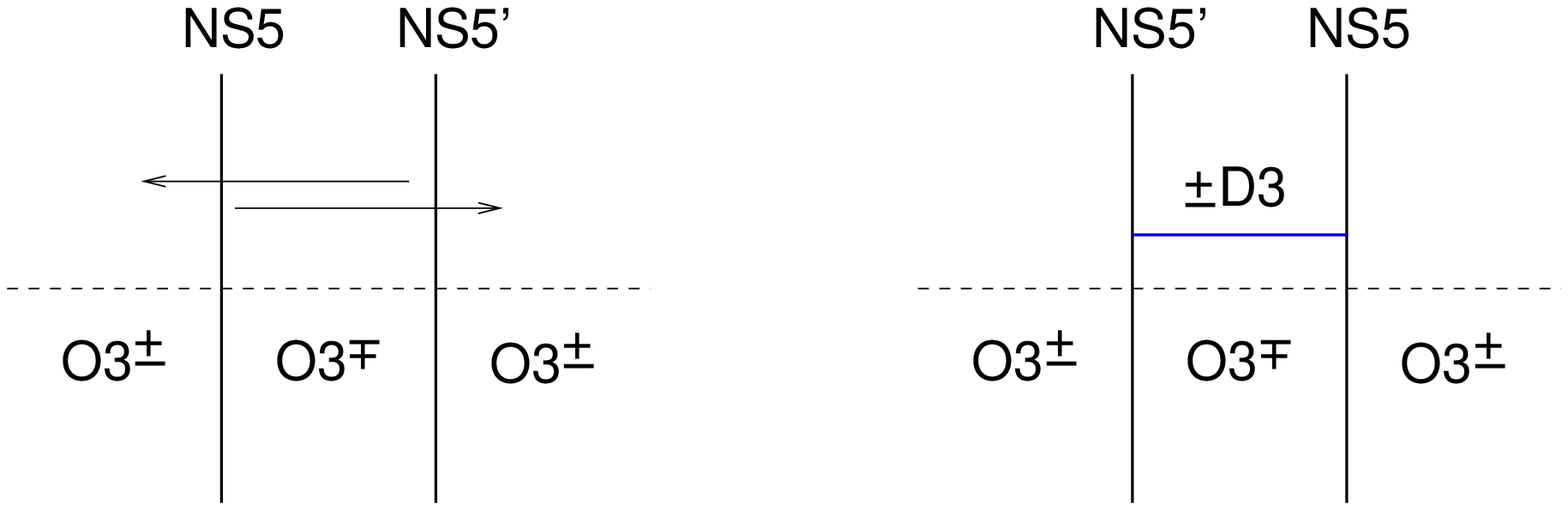}}

As in the case without the orientifold, we expect the ``s-rule''
and the brane creation effect to lead to a restriction on the
number of inequivalent superconformal theories. However one has to
be slightly careful here since both the brane creation effect, and
the ``s-rule'' which can be derived from it, are shifted in the
presence of an O3-plane relative to the case without the
orientifold. The basic reason for the shift is an additional brane
creation effect when two purely NS5-branes, that intersect an
O3-plane, cross. In particular, if the orientifold between the
NS5-branes is an O$3^-$-plane a single D3-brane is created, and if
it is an O$3^+$-plane a single D3-brane is annihilated (see
\newbranecreation ). If the orientifold between the NS5-branes is
an $\widetilde{{\rm O} 3}^-$ or $\widetilde{{\rm O} 3}^+$-plane
there is no brane creation. The creation (or annihilation), or
lack thereof, can be understood from the requirement that the jump
in the D3-brane charge across each NS5-brane is preserved when the
NS5-branes cross.\foot{One can derive this
effect from a linking number analogous to the NS5-D5 case
\HananyIE . The creation of branes by NS5-branes crossing on an
orientifold plane was originally proposed in type IIA brane
configurations for four-dimensional ${\cal N}=1$ gauge theories
with orthogonal and symplectic gauge groups in \Shapere , in order
to reproduce Seiberg duality for these theories \IntriligatorID .
In that case the jump in the D4-brane charge across an NS5-brane
leads to an asymptotic bending of the NS5-brane, which must be
preserved in finite (zero) energy brane motions.} The total number
of D3-branes created by the crossing NS5-brane and $(1,2k)$5-brane
therefore depends on the type of O3-plane between them (on the
side that initially shrinks). In the O$3^-$ case $k+1$ D3-branes
are created, in the O$3^+$ case $k-1$ are created, and in the
$\widetilde{{\rm O} 3}^-$ and $\widetilde{{\rm O} 3}^+$ cases $k$
D3-branes are created. This also implies that the restriction on $l$ is
shifted to $l\leq k+1$ in the first case (class I), $l\leq k-1$ in
the second case (class II), and remains $l\leq k$ in the last two
cases (classes III and IV).

\ifig\braneequivalences{The brane description of the parity-dualities.}
{\epsfxsize4in\epsfbox{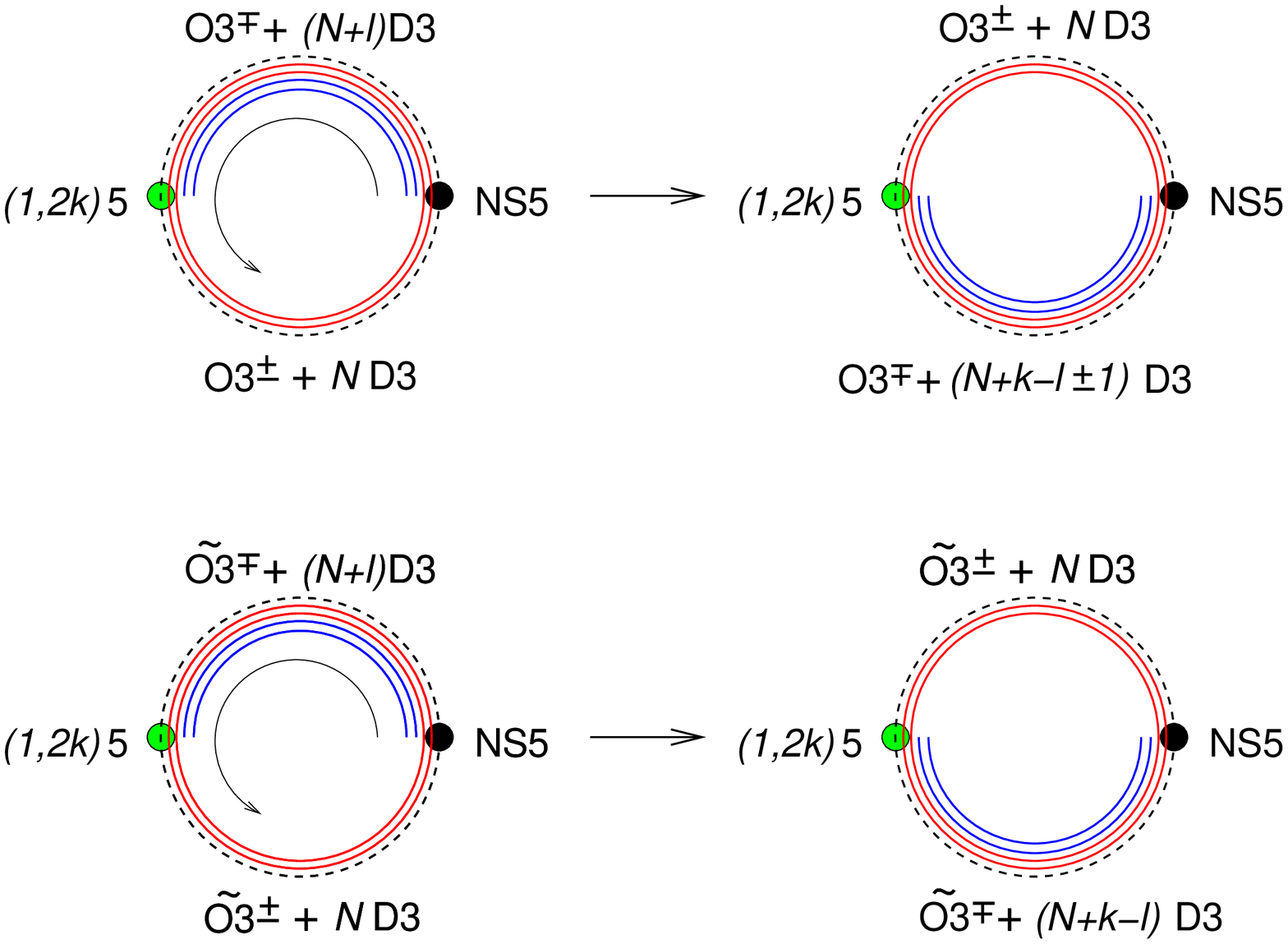}}

We are now ready to examine the effect of moving one of the 5-branes
around the circle. As before, we find that only one motion gives rise to another
supersymmetric configuration, and therefore to another ${\cal N}=3$
supersymmetric gauge theory,
namely the one where the 5-brane winds the circle once
in the direction that initially shrinks the suspended D3-branes.
The initial configuration has $l$ suspended D3-branes in one of the
segments.
The final configuration has the two orientifold segments exchanged,
so $k\rightarrow -k$,
and $k-l$ D3-branes, with the possible $\pm 1$ shift, in the other segment.
The resulting pairs of brane configurations
for the four classes of theories are shown in \braneequivalences .
This leads us to conjecture the following parity-dualities between
the corresponding superconformal field theories:
\eqn\parityduality{
\eqalign{
I_k(l) & =  I_{-k}(k-l+1), \cr
II_k(l) & =  II_{-k}(k-l-1), \cr
III_k(l) & =  III_{-k}(k-l), \cr
IV_k(l) & =  IV_{-k}(k-l) \,.
}}
This identification
is consistent with the bounds described above:
\eqn\maxl{
0 \leq l_I \leq k+1 \,,\quad
0 \leq l_{II} \leq k-1 \,,\quad
0 \leq l_{III,IV} \leq k \,.
}
Looking at \fourclasses\ we also see that at the extremal values the parity-dualities
of \parityduality\ give equivalence relations that map between the different classes:
\eqn\equivalentclasses{
\eqalign{
I_k(k+1) & = II_k(0), \cr
II_k(k-1) & = I_k(0), \cr
III_k(k) & = IV_k(0), \cr
IV_k(k)&  = III_k(0) \,.
}}
All in all, we therefore have $4k$ inequivalent theories (for each value of $N$ and $k$).

As in the case with the unitary gauge groups, the two theories
related by a parity-duality are never weakly coupled at the same time, making it
difficult to test this duality.
However we will provide more evidence for these dualities, as well as for the
maximal values of $l$, by analyzing the superconformal theories directly,
and we will also show that this picture is consistent with the gravitational dual
descriptions.

\subsec{The $O(M)\times USp(2N)$ superconformal theories}

The three-dimensional orthogonal/symplectic ${\cal N}=3$ gauge theories
living on the brane
configurations described above flow in the IR to superconformal
Chern-Simons-matter theories.
Alternatively we can also get these superconformal theories by gauging
a discrete symmetry of the ${\cal N}=6$ superconformal theories
with
$U({\tilde M})_{2k}\times U({\tilde N})_{-2k}$
which includes the usual orientifold action on the gauge fields, and the action
on the matter fields in \orientifoldmatter .
This action breaks the baryon number symmetry $U(1)_b$ completely,
while the subgroup
of the R-symmetry $SU(4)_R$ left unbroken by $M_{IJ}$ is $USp(4)\sim SO(5)$.
The new
$O(M)\times USp(2N)$ gauge
theory must therefore have ${\cal N}=5$ supersymmetry, and this was verified
(classically) in \Hosomichi
\foot{Note that one could also consider ${\cal N}=5$ supersymmetric
theories with the gauge group $SO(M)\times USp(2N)$, but we do not obtain these
theories from our brane construction or from the projection of the unitary theories,
and we will not consider them in this paper.
For the special case of $SO(2)\times USp(2N)$ these theories have an enhanced
${\cal N}=6$ supersymmetry \refs{\Hosomichi,\SchnablWJ}.}.

Let us now determine the moduli space of these theories.
A complete analysis of the moduli space of these theories is given in the appendix.
Here we will motivate the result by a shortcut, starting
with the moduli space of the parent ${\cal N}=6$ superconformal theory.
The latter is given by
$({\bf C}^4/{\bf Z}_{2k})^{\tilde N}/{\bf S}_{\tilde N}$,
where the ${\bf Z}_{2k}$ is generated by the transformation
$C_I \rightarrow e^{\pi i/k} C_I$.
It is easy to see that the symmetry group of \orientifoldmatter\ is ${\bf Z}_4$.
This symmetry does not commute with the ${\bf Z}_{2k}$ symmetry acting on the moduli
space of the original
theory. In fact, the square of the ${\bf Z}_4$ generator acts trivially on the gauge
fields, and on the matter fields it gives $C_I \rightarrow -C_I$, which corresponds
precisely to the generator of the ${\bf Z}_{2k}$ symmetry raised to the $k$'th power.
This is precisely the definition of the binary dihedral group $\widehat{\bf D}_{k}$
(we are using the convention that $\widehat{\bf D}_k$ is the binary dihedral group
with $4k$ elements).
The simplest one is $\widehat{\bf D}_1={\bf Z}_4$, and the higher $k$
groups are non-Abelian. Thus, we expect the moduli space after the projection to be
$({\bf C}^4/\widehat{\bf D}_k)^N/S_N$, and we verify in the appendix that this is
indeed the case.

Based on this moduli space, it is natural to conjecture that our theories are related
to the theory of $N$ M2-branes
probing the orbifold ${\bf C}^4/\widehat{\bf D}_k$ in M theory. This orbifold preserves
precisely ${\cal N}=5$ supersymmetry \Morrison, which is consistent with this conjecture.
This conjecture implies that the dual M theory
background is $AdS_4\times S^7/\widehat{\bf D}_k$.
We also expect a regime of the parameters $N$ and $k$ in which there
is a weakly curved type IIA string theory description.
We will discuss both gravitational duals below.

But first, let us return to the implications of the ``s-rule'' and brane creation effect,
namely the parity-duality relations \parityduality , and the maximal values
of $l$ \maxl .
We will use similar arguments to the ones we used for the $U(N+l)\times U(N)$
theories as further evidence for these quantum properties of the orientifold
theories.

We start by considering a generic point on the moduli space.
This leaves at low energies a pure ${\cal N}=3$ supersymmetric
Chern-Simons theory with a gauge group that depends
on $l$. We can then integrate out the gauginos to get an effective bosonic Chern-Simons
theory with a shifted level, where
the shift is given by the dual Coxeter number
of the gauge group, $k\rightarrow k - h\cdot {\rm sign}(k)$ \WittenDS \foot{In
\WittenDS\ the shift was by $-{h\over 2}\cdot {\rm sign}(k)$ for the ${\cal N}=1$ theory with one gaugino.
In the ${\cal N}=3$ theory there are four gauginos, three of which contribute
with the same sign, and the fourth with the opposite sign.}.
The dual Coxeter numbers of the groups $O(2l)$, $O(2l+1)$ and $USp(2l)$
are given, respectively, by $h=2l-2$, $2l-1$ and $l+1$.
We can then determine the maximal value of $l$, as in the previous section, by requiring that the level in the effective bosonic CS theory should have the same sign as the original level,
so that the one-loop shift of the level in this theory will bring us back
to the original level $k$. This gives $0\leq l_I\leq k+1$, $0\leq l_{II}\leq k-1$ and
$0\leq l_{III}\leq k$, for the class I, II and III theories, respectively,
precisely as predicted above.

The class IV theories are a bit more subtle, because on their moduli space we obtain
a $USp(2l_{IV})\times O(1)$ gauge theory; the $O(1)$ theory does not have any gauge fields,
but it leads to massless matter fields in the fundamental of $USp(2l_{IV})$. Thus, we
cannot map these theories directly to bosonic CS theories. We believe that these
superconformal theories exist precisely for $0 \leq l_{IV} \leq k$, and it would be interesting
to verify this directly.


Our main direct evidence for the parity-dualities in \parityduality\ comes (as in the unitary case)
from level-rank duality between the two
effective bosonic Chern-Simons theories on the moduli space.
For example, the $I_k(l)$ theory gives at a generic point on the moduli space
an ${\cal N}=3$ $O(2l)_{2k}$
Chern-Simons theory, which is equivalent to a bosonic $O(2l)_{2k-2l+2}$ Chern-Simons
theory, while the $I_{-k}(k-l+1)$ theory on the moduli space includes an
${\cal N}=3$ $O(2k-2l+2)_{-2k}$ theory, which is equivalent to a bosonic
$O(2k-2l+2)_{-2l}$ theory. The two WZW theories that we obtain from these in the presence
of boundaries are related by level-rank duality,
which supports the first of the four conjectured dualities in \parityduality .
A similar test of the $II_k(l)$ and $III_k(l)$ dualities confirms their consistency as well.
For the class $IV$ theories it is again more difficult to analyze this directly because
of the presence of extra massless fields, so we do not have direct field theory arguments
for the duality in this case.

\subsec{M theory lift}

The configuration of D3-branes, O3-planes,
and fivebranes described above can be T-dualized and lifted to
M theory as in \ABJM, resulting in M2-branes probing a singularity in a toric
hyperK\"ahler 8-manifold, which preserves $3/16$ of the
supersymmetry \kkmonopole. This geometry has a $T^2$ formed from the
T-dualized circle and the M theory circle, fibred over a six
dimensional base. The IR limit of the field theory on the
D3-branes that we derived corresponds to taking the size of the
wrapped circle small in IIB, and hence large in the T-dual
picture. Further, the low energy limit of the D2-branes brings us
to strong coupling, so the superconformal field theory describes
the limit of large $T^2$ in the hyperK\"ahler geometry.

In \ABJM, the hyperK\"ahler geometry resulting from the theory with one
NS5-brane and one $(1,k)$ 5-brane was described explicitly, and
its IR limit was shown to be a ${\bf C}^4/{\bf Z}_k$ singularity.
The hyperK\"ahler geometries were studied in \kkmonopole,
 and have a
general form related to a product of two Taub-NUT spaces,
\eqn\formofm{\eqalign{
  ds^2 = & U_{ij} d \vec x^i \cdot d \vec x^j + U^{ij} ( d \varphi_i + A_i ) ( d \varphi_j  + A_j)~,
 \cr
 A_i = & d \vec x^j \cdot \vec \omega_{ji }  = dx^j_a \omega^a_{ji} ~,~~~
 ~~ ~~~ \partial_{ x_a^j} \omega^{b}_{k i } - \partial_{x_b^k} \omega^a_{ji} =
  \epsilon^{ab c} \partial_{x^j_c} U_{ki}~,}}
  where $U_{i j}$ is a two by two symmetric matrix of harmonic
 functions
and $U^{ij}$ is its inverse.
In our case there will be an additional ${\bf Z}_2$ quotient due to the
lift of the orientifold.

Suppose we first consider the T-dual and lift of the O3-plane without
any 5-branes. Then, in M theory the geometry would simply be a
${\bf Z}_2$ quotient\foot{More precisely, we obtain four such
quotients on four different points on the $T^2$, but we will focus
here on the vicinity of one of these four orbifold points.},
acting on the coordinates above as
$\vec{x}_i\rightarrow - \vec{x}_i$, and $\varphi_i \rightarrow -\varphi_i$,
where $\vec{x}_i$ describe a flat ${\bf R} ^6$, and $\varphi_i$
describes the fiber $T^2$. Re-introducing the fivebranes, the
torus fiber will shrink along 3-planes in the base. However, far
from these centers of the Taub-NUT spaces, the geometry is not altered
significantly, and one sees that the ${\bf Z}_2$ action must be
the same as above.

The particular geometry obtained from an NS5-brane and a $(1,2k)$
5-brane is described by
\eqn\twokk{
 U= {\bf 1} +   \pmatrix{ h_1 & 0 \cr 0 & 0 } +
 \pmatrix { h_2  & 2 k h_2 \cr 2 k h_2 & 4 k^2 h_2 },   ~~~~~
  h_2 = { 1 \over  2 |\vec x_1 + 2 k \vec x_2 | } ~,~~~~h_1 = { 1 \over 2 | \vec x_1 |
  }~,
  } in which the singularity at the origin is ${\bf C}^4/{\bf
  Z}_{2k}$, as explained in \ABJM. Note that near the origin of
the hyperK\"ahler 8-manifold, one obtains ${\bf C}^4/{\bf Z}_{2k}$
using the fact that the center of
each Taub-NUT geometry looks just like a
flat ${\bf R}^4$ space, naturally written in polar coordinates with a
Hopf-fibred angular $S^3$.

More explicitly, the Hopf fibration defines a map $f: {\bf C}^2
\rightarrow S^2$ which can be written as
$f(z_1, z_2) = \left(2 {\rm Re}(z_1 z_2^*), 2 {\rm Im}(z_1
z_2^*), |z_1|^2 - |z_2|^2 \right)$, where the $S^2$ is regarded as
the unit sphere in ${\bf R}^3$, the base of the Taub-NUT. The
orientifold ${\bf Z}_2$ then acts as the antipodal map on the
$S^2$, which can be seen to lift to the action $z_1 \rightarrow i
z_2^*, \ z_2 \rightarrow - i z_1^*$ on ${\bf C}^2$, noting that
the overall phase of $z_1, z_2$ is exactly the Hopf circle, which
is also inverted, $\varphi_1 \rightarrow -\varphi_1$, by the lift
of the orientifold. This ${\bf Z}_2$ also acts simultaneously on
the other ${\bf C}^2$ in the eight dimensional geometry.

Therefore we conclude that the transverse geometry to the M2-branes
includes a ${\bf C}^4/\widehat{\bf D}_k$ singularity (which will control the
IR field theory). The ${\bf Z}_{2k}$ subgroup of $\widehat{\bf D}_k$ is the
orbifold obtained from the 5-branes, acting as $(z_1, z_2, z_3,
z_4) \rightarrow e^{\pi i \over k} (z_1, z_2, z_3, z_4)$, while the
other generator acts by
\eqn\ztwoaction{(z_1, z_2, z_3, z_4) \rightarrow (i
z_2^*, - i z_1^*, i z_4^*, - i z_3^*).}
The latter is a ${\bf Z}_2$
transformation on the ${\bf C}^4/{\bf Z}_{2k}$ orbifold, but of course it gives a
${\bf Z}_4$ in the ${\bf C}^4$ covering space.

It is not hard to check that this is identical to the ${\bf
C}^4/\widehat{\bf D}_k$ singularity mentioned in \Morrison, in different
coordinates. One sees that the $\widehat{\bf D}_k$ sits inside an $SU(2)$
subgroup of the $SO(8)$ rotation group of ${\bf R}^8$, and thus its commutant is the
$SO(5)_R$ R-symmetry of this ${\cal N}=5$ orbifold.

In the special case of $k=1$, the orbifold is simply ${\bf C}^4/{\bf Z}_4$,
which is unique up to coordinate redefinitions.
Therefore the $O\times USp$ theories actually preserve ${\cal N}=6$ supersymmetry
for $k=1$, and they should be quantum mechanically dual to the ${\cal N}=6$
Chern-Simons-matter theories with unitary gauge groups (with the same minimal
rank) at level $k=4$.
In particular, the four $O\times USp$ theories in this case are
$O(2N)_2\times USp(2N)_{-1}$, $O(2N+2)_2\times USp(2N)_{-1}$,
$O(2N+1)_{2}\times USp(2N)_{-1}$, and $USp(2N)_1\times O(2N+1)_{-2}$.
The first two should then presumably be equivalent to the $U(N)_4\times U(N)_{-4}$
and $U(N+2)_4\times U(N)_{-4}$ theories (though we cannot say which to which),
and the last two to $U(N+1)_4\times U(N)_{-4}$ and $U(N+3)_4\times U(N)_{-4}$.
Of course, these theories are strongly coupled, so it is
difficult to check these identifications, but one can compare their spectrum of
chiral operators, their moduli spaces, and so on.
As a simple consistency check, note that the parity-dualities \parityduality,
\equivalentclasses\ imply that the first two $O\times USp$ theories mentioned above
map to themselves by a parity transformation. This is obviously true also for
the $U(N)_4\times U(N)_{-4}$ theory, and it turns out to also be true for the
$U(N+2)_4\times U(N)_{-4}$ theory, using the equivalences \equivalence. Similarly,
our last two $O\times USp$ theories map to each other under parity, and using \equivalence\
we can check that the same is true for the $U(N+1)_4\times U(N)_{-4}$ and
$U(N+3)_4\times U(N)_{-4}$ theories.

\subsec{M theory dual}

From the M theory lift above, it is easy to construct the M theory background
dual to our orientifolded field theories \fourclasses. This background is simply
$AdS_4\times S^7 / {\widehat {\bf D}}_k$, with $N$ units
of 4-form flux in $AdS_4$:
\eqn\orientM{ \eqalign{ ds^2 =&  { R^2
\over 4 } ds^2_{AdS_4} + R^2 ds^2_{S^7/{\hat D}_k}, \cr G_4 \sim &
N \epsilon_4, }}
where $\epsilon_4$ is the unit volume form on
$AdS_4$. This is obviously consistent with the moduli space of these
field theories which we described above. The spectrum of chiral operators in this
background was recently computed in \HananyQC. The eleven dimensional
supergravity approximation is valid for $k\ll N^{1/5}$, as before.

It is natural to describe the $S^7/{\widehat {\bf D}}_k$ as a ${\bf
Z}_2$ quotient of the $S^1$ fibred over ${\bf CP}^3$ that we had
in the $S^7/{\bf Z}_{2k}$ case. This is equivalent to a
twisted circle bundle over the non-orientable 6-manifold ${\bf
CP}^3/{\bf Z}_2$ (where by twisted we mean that the curvature
of the circle bundle lives in the ${\bf Z}_2$-twisted cohomology).

The geometry described above has a ${\bf Z}_{4k}$ torsion 3-cycle; generally,
the discrete torsion in M theory when we divide by a discrete group with $|\Gamma|$
elements is ${\bf Z}_{|\Gamma|}$ \Sethi. One way to describe this cycle explicitly
is as the twisted $S^1$ fibration over an ${\bf
RP}^2$ in the base. This cycle is precisely half the torsion cycle
in $S^7/{\bf Z}_{2k}$, given by the $S^1$ fibred over a ${\bf
CP}^1$. In M theory, one can thus turn on a discrete torsion $G_4$
flux, or in other words, a flat $C_3$ field with holonomy in ${\bf
Z}_{4k} \subset U(1)$ on that torsion 3-cycle. Explicitly,
\eqn\Cflux{C_3 \propto {l \over 4k} J \wedge d\varphi,}
in the $S^7/{\widehat {\bf D}}_{k}$, with $l=0,\cdots,4k-1$; this is invariant under the identifications
described above. Thus, we have $4k$ different field theories, just as expected
from our discussion above. Upon
reduction to type IIA string theory, we will see precisely how this discrete torsion matches
with the $4k$ orientifold field theory duals found earlier.

\subsec{Type IIA string theory orientifold dual}

When $N \ll k^5$ the M theory circle becomes small, and it is more appropriate to
describe the gravitational dual using type IIA string theory (which is weakly curved
for $N \gg k$). The reduction to type IIA
is very similar to the one discussed in the previous section, \twoasol\ (with $2k$
replacing $k$ everywhere). The only difference is the additional ${\bf Z}_2$ identification
which we have to take into account. This identification acts on the natural projective
coordinates of ${\bf CP}^3$ by \ztwoaction. It is easy to check that this maps $J \to -J$,
so that the orientation of ${\bf CP}^3$ (and, thus, of the full type IIA background) is
reversed. Thus, our type IIA background is an orientifold of $AdS_4\times {\bf CP}^3$
(the orientifold has no fixed points so this is a smooth manifold).
In M theory, the ${\bf Z}_2$ acts on the ${\bf CP}^3$ as above, together with an inversion of
the coordinate on
the M theory circle (while not acting on the 3-form field). Thus, in type IIA string
theory, beyond its geometric action described above, the orientifold inverts the RR 1-form
$C_1$ and the NSNS 2-form $B_2$, while leaving invariant the RR 3-form $C_3$. Note that
since $F_2 \to -F_2$ and $J \to -J$, this is consistent with the RR 2-form flux $F_2 = 2 k J$
(which is invariant under the ${\bf Z}_2$). One can think of this as $k$ units of RR 2-form
flux on the cycle ${\bf CP}^1/{\bf Z}_2$ that we obtain by performing the identification
\ztwoaction\ on the ${\bf CP}^1 \subset {\bf CP}^3$. Note that this identification maps the
${\bf CP}^1$ (defined, say, by $z_3=z_4=0$) to itself, so that we obtain a smaller 2-cycle,
but it does not map the ${\bf CP}^2$ cycle (defined, say, by $z_4=0$) to itself. Thus,
the minimal 4-cycle in the orientifolded background is still ${\bf CP}^2$.

Both $U(1)$ gauge fields which are present on
$AdS_4\times {\bf CP}^3$ before the orientifold, are projected out by the orientifold
(one of them originally became
massive by swallowing an axion \ABJM , but this axion is also projected out by the orientifold).
The projection on the $SU(4)$ gauge fields leaves an $SO(5)$ subgroup invariant, so that
the gauge group on $AdS_4$ agrees with the global symmetry group of our field theories.

Next, let us analyze the
reduction of
discrete torsion in this picture. The fluxes $f_4$ and $b_2$
described in the previous section are invariant under the ${\bf Z}_2$. As in the previous
section we can define
$f_4 \propto \int_{{\bf CP}^2} {\tilde F}_4$ (normalized to be an integer in the
absence of 2-form flux), but we can now define a
smaller 2-form holonomy
${\tilde b}_2 = {1\over {2\pi}} \int_{{\bf CP}^1/{\bf Z}_2} B_2$ (which is periodic with
period one). The same arguments as in the previous
section imply that the
quantized four-form flux is $f_4 + 4 k {\tilde b}_2$ (since we have $2k$ units of $F_2$
flux on ${\bf CP}^1$, and ${\tilde b}_2 = {1\over 2} b_2$).
Thus, we have $4k$ possibilities for the NSNS holonomy (when ${\tilde F}_4=0$), given by
${\tilde b}_2 = l/4k$ with $l=0,\cdots,4k-1$
(up to a possible integer shift).
This precisely agrees with our M theory expressions in the
previous subsection.

As in the previous section, the fact that the different $I_k(l)$ theories are related
by adding fractional branes (each of which shifts ${\tilde b}_2$ by ${1\over {2k}}$) makes it
natural to identify the $I_k(l)$ theory with the type IIA background with
$\tilde{b}_2 = {l\over 2k}$. However, the fractional brane picture only tells us the
differences between the values of $\tilde{b}_2$ between the different theories, and not
their absolute values, so it allows for an identification of the form $\tilde{b}_2 =
{{l+c}\over {2k}}$ with some (half)-integer $c$.
The $III_k(l)$ theory is then related to this by adding
another ``half of a fractional brane'' (in the brane construction), so it is natural
to identify it with the background with ${\tilde b}_2 = {{2l+2c+1}\over {4k}}$. Using the
equivalence relations \equivalentclasses, we can then determine the identifications of the
class II and class IV theories. Our full identifications are thus summarized by:
\eqn\bcorrespondence{
\eqalign{
I_k(l) & \leftrightarrow  {\tilde b}_2 = {l+c \over 2k}, \cr
II_k(l) & \leftrightarrow {\tilde b}_2 = {l+c + k + 1\over 2k}, \cr
III_k(l) & \leftrightarrow {\tilde b}_2 = {l+c + {1\over 2}\over 2k}, \cr
IV_k(l)&  \leftrightarrow {\tilde b}_2 = {l+c + k + {1\over 2}\over 2k} \,.
}}
For any value of $c$ we have a one-to-one map between our space of theories and
the space of type IIA backgrounds.
In order to determine $c$, we require consistency with
the parity-duality identifications of \parityduality.
Recall that the parity transformation
of type IIA
takes ${\tilde b}_2 \to -{\tilde b}_2 =
1 - {\tilde b}_2$, and we conjectured that it acts on our theory space as \parityduality.
All of these identifications turn out to be consistent for precisely two values of $c$ :
$c = -{k+1\over 2}$ or $c = {k-1\over 2}$ (mod $2k$).
Note that the NSNS holonomy encodes in the language of our original
type IIB brane construction both the number of fractional branes and the NSNS
and RR torsion fluxes, as a result of the chain of dualities that we
performed.\foot{The type
IIA orientifold
may
also admit RR torsion corresponding to a discrete holonomy of the
RR 3-form $C_3$ on a torsion 3-cycle in ${\bf CP}^3/{\bf Z}_2$
($H_3({\bf CP}^3/{\bf Z}_2,{\bf Z})={\bf Z}_2$).
Since we have no room for an extra quantum number associated with this,
we expect that this torsion should be correlated with the parity of $4 k \tilde{b}_2$,
that distinguishes the class I and II theories from the class III and IV theories (and
is related to the RR torsion in our brane construction).
This correlation may be seen in M theory by noting that the torsion 3-cycle
lifts to the same torsion cycle in M theory as ${\bf CP}^1/{\bf Z}_2$.
It would be interesting to
derive this correlation directly in type IIA string theory by a careful analysis
of the flux quantization conditions in our background.}

It is not clear how to determine which of the two possibilities for $c$ is correct. Note that for both possibilities,
the $I_k(0)$ theory does not map to ${\tilde b}_2=0$ as one might have naively expected; presumably
the orientifold itself induces some
discrete torsion in our background,
even in the absence
of any fractional branes. This is consistent with the fact that the $I_k(0)$ theory is not
parity-invariant
(for $k > 1$).
It would be interesting to understand this better, and to determine
the precise value of $c$.
Since we have no massless $U(1)$ fields, the di-baryon operators
(coming from D4-branes wrapped on ${\bf CP}^2$) are not independent of the ``meson''-type
operators, in agreement with our gauge theory expectations. Thus, these operators do not seem
to give any clues about the correct identifications.

It would be interesting to analyze the mass deformations of these theories (analogous
to \GomisVC) and to try to obtain a correct count of the number of vacua in these theories.
It would also be interesting to see if analyzing the operators in this theory
in the 't Hooft large $N$ limit
leads to
interesting integrable spin chains, either on the field theory or on the string theory
side. Of course, at the leading (planar) order the theories of this section are identical
to those of the previous section. The leading difference between them comes from
non-orientable diagrams at the next order in $1/N$.

\vskip 1cm

{\centerline {\bf Acknowledgments }}

We would like to thank Juan Maldacena for collaboration on the first part of
this paper, and for many useful discussions and comments.
We would like to thank John Schwarz for comments on an early version of the paper.
We would also like to thank
Davide Gaiotto, Sergei Gukov, Dario Martelli, Alfred Shapere,
Yuji Tachikawa, Mark Van Raamsdonk, and Edward Witten for useful discussions.
The work of OA was
supported in part by the Israel-U.S. Binational Science
Foundation, by a center of excellence supported by the Israel
Science Foundation (grant number 1468/06), by a grant (DIP H52) of
the German Israel Project Cooperation, by the European network
MRTN-CT-2004-512194, and by Minerva. OA would like to thank the ``Monsoon workshop
on string theory'' at Mumbai, India, for hospitality during the course of this work.
OB gratefully acknowledges support from the Institute for Advanced Study.
The work of OB was
supported in part by the Israel Science Foundation under grant
no.~568/05. The work of DJ was supported in part by DOE grant
DE-FG02-96ER40949. DJ would like to thank the Simons Center for Geometry and Physics at Stony
Brook and the 2008 Simons Workshop in Mathematics and Physics for kind
hospitality during the course of this work.

\appendix{A}{Moduli spaces of the orientifold field theories}

In this appendix we compute the classical moduli spaces of the
$O(2M)_{2k} \times USp(2N)_{-k}$ and $O(2M+1)_{2k}\times USp(2N)_{-k}$
field theories discussed above; these are
expected to be exact quantum mechanically due to the considerable
amount of supersymmetry. Recall that the moduli space of the $U(N)
\times U(M)$ Chern-Simons-matter theory consists of diagonal
matrices \ABJM, so one might expect the same here. However, the
orientifold projection \orientifoldmatter\ does not preserve the
diagonal form due to the presence of $J$. This is reflected in the
fact that it is impossible to use the projected $O \times USp$
gauge symmetry to diagonalize a generic matrix in the
bi-fundamental, in contrast to the unitary group case.

We will choose $J$ to be made of $2\times 2$ blocks in which it is
equal to $(-i\sigma_2)$.
Using the gauge symmetry, it is then possible to bring the vacuum
expectation value
of one of the complex bi-fundamental matter fields, $A_1$, into a $2 \times 2$ block
diagonal form. It is easy to check that on the moduli space $A_2$
must also have this form. Within each block we then just need
to determine the moduli space of the $O(2)\times USp(2)$ theory. In
the first part of this appendix we show that the moduli space of this
theory is ${\bf C}^4/\widehat{{\bf D}}_k$, on which the gauge group is
generically broken to
a $U(1)$, in agreement with our interpretation of this theory as
describing a single
M2-brane in that geometry; moreover, that is the result expected by
projecting the moduli space of the $U(2N) \times U(2N)$ theory
found in \ABJM.

Therefore, the moduli space of the $O(M)_{2k} \times USp(2N)_{-k}$
theories for $M \geq 2N$ is given by $N$ copies of the $O(2)
\times USp(2)$ moduli space, quotiented by the unbroken permutation
symmetry $S_N \subset O(2N)
\times USp(2N)$, namely it is equal to $Sym^N\left( {\bf
C}^4/\widehat{{\bf D}}_k \right)$. The low-energy effective theory on
this moduli space includes a residual
pure ${\cal N}=3$ Chern-Simons theory $O(M-2N)_{2k}$.

For the $USp(2M)_k \times O(2 N)_{-2k}$ theories with $M \geq N$
the moduli space is similarly given by $Sym^N\left({\bf
C}^4/\widehat{\bf D}_k \right)$, where now the low-energy effective
 action on the moduli space includes an ${\cal N}=3$ supersymmetric
$USp(2M-2N)_k$ pure Chern-Simons theory. The situation for the
$USp(2M)_k \times O(2N+1)_{-2k}$ theories is similar, except that the
low-energy effective theory on the moduli space now includes an ${\cal N}=3$ supersymmetric
Chern-Simons-matter theory with gauge group $USp(2M-2N)_k \times O(1)_{-2k}$, which has
massless matter in addition to the ${\cal N}=3$ Chern-Simons terms. In the
second part of this appendix we
show that there are no additional moduli associated to this
residual theory, so it does not affect the moduli space.

\subsec{The moduli space of the $O(2)_{2k} \times USp(2)_{-k}$
theory}

We begin by gauge fixing the gauge fields to zero, so that the moduli space
is given by the zero locus of the bosonic potential for the two bi-fundamental
scalar fields $A_1, A_2$
(which we write as $2\times 2$ matrices).
This
potential can be written as a sum of squares, coming both from the usual
F-terms $|\partial W|^2$ (with $W$ given by \orientsuperpot), and from the
$\sigma$-terms in the supersymmetric kinetic terms. These terms
are proportional to
\eqn\Dtermeqn{
\sum_{a=1}^2 \tr(\left|
\sigma_{O} A_a - A_a \sigma_{USp} \right|^2),}
where the scalar fields $\sigma$
in the vector multiplet are
equal (up to a constant that we will ignore)
to the moment maps,
\eqn\momentmaps{\sigma_O = \sum_{a=1}^2 (A_a A_a^{\dag} - A_a^* A_a^T), \qquad
\sigma_{USp} = \sum_{a=1}^2 (A_a^{\dag} A_a + J A_a^T A_a^* J),}
after integrating
out the auxiliary fields in the vector multiplets. Their form can either be
derived directly, or by using the projection from the unitary case.

It is easy to see that both the F-term equations
\eqn\Ftermcond{A_1^T A_2 J A_2^T = A_2^T A_2 J A_1^T,\qquad A_2^T A_1 J A_1^T = A_1^T A_1 J A_2^T,}
and the equations coming
from \Dtermeqn\ are satisfied when $[J, A_1] = [J, A_2]=0$, since in that
case the matrices take the form
\eqn\matrixansatz{A_1 = x + y J, \ \ \ \ \ A_2=z +w J,}
so that the matrices (and their conjugates and transposes) all commute with
each other.
The scalar fields in the vector multiplets are then equal to
\eqn\sigmas{\sigma_{O(2)} = 2(y x^*-x y^* +w z^*- z w^*) J =
\sigma_{USp(2)},}
so it is easy to check that \Dtermeqn\ vanishes.
%
Therefore, the matrices \matrixansatz\ are on the moduli space for any
complex numbers $x, y, z, w$.

Generically, on the moduli space, the gauge symmetry is broken to a diagonal
$U(1)$, acting by an $O(2)$ transformation on the left, and by the
inverse transformation (which is contained in $USp(2)$) on the
right. The unbroken gauge field is thus ${\cal A}_+ = {\cal
A}_{O(2)} + {\cal A}_2$, where we define ${\cal A}_{USp(2)} = \sum_{j=1}^3
\sigma_j {\cal A}_j$. The other three generators of the gauge
group become massive. However, the off-diagonal combination ${\cal
A}_- = {\cal A}_{O(2)} - {\cal A}_2$ acts on the moduli space,
that is, it preserves the form \matrixansatz. This rotation acts
on the components as
\eqn\Uoneaction{\pmatrix{x \cr y} \rightarrow
 \pmatrix{\cos \phi & -\sin\phi \cr \sin\phi & \cos\phi}
\pmatrix{x \cr y},}
and similarly for $z$ and $w$.
It is convenient to define new coordinates on the moduli space $u_1 = {\rm Re}(x) + i
{\rm Re}(y)$, $u_2 = {\rm Im}(x) + i {\rm Im}(y)$, $u_3 = {\rm Re}(z) + i
{\rm Re}(w)$, $u_4 = {\rm Im}(z) + i {\rm Im}(w)$, so
that this broken $U(1)$ acts by multiplying all the $u_I$ ($I=1,2,3,4$) by
$e^{i\phi}$.

Next, we need to take into account the effect of constant gauge transformations
(which leave the gauge fields vanishing) acting on the moduli space.
Since effectively we have a $U(1)\times U(1)$ theory, this analysis is precisely
the same as in \ABJM. The only difference is that the level of the $U(1)$ Chern-Simons terms is
now $2k$ instead of $k$ as in \ABJM; this is clear for the $O(2)_{2k}$ piece, and
one can show that it is also true for the $U(1)$ in $USp(2)_{-k}$ by considering the
embedding of $U(1)$ in $USp(2)$. Thus, by a similar analysis to \ABJM\ (either by
considering which gauge transformations leave the CS terms invariant, or by dualizing
the gauge field ${\cal A}_+$ to a scalar), we find that the $U(1)$ action leads to
a ${\bf Z}_{2k}$ identification, where the ${\bf Z}_{2k}$ acts as
$u_I \to \exp(\pi i n / k) u_I$ (for $n \in {\bf Z}$).

This takes into account the connected part of the gauge group.
However, $O(2)$ has an additional disconnected component (involving
matrices with determinant $(-1)$), and
this leads to an additional ${\bf Z}_2$
identification of the moduli space.
In particular, we can consider a gauge transformation
\eqn\Ztwoquot{
A_1 \rightarrow
\pmatrix{ 0 &1 \cr 1 & 0} A_1 \pmatrix{i & 0 \cr 0 & -i} =
\pmatrix{-i y & -i x \cr ix & -i y},}
(and similarly for $A_2$). This
preserves the form
\matrixansatz, and acts on the coordinates defined above as
$(u_1,u_2, u_3, u_4) \rightarrow (i u_2^*, -i u_1^*, i u_4^*, -i
u_3^*)$. This is exactly the orientifold action we discussed
in section 3,
which squares to
$u_I \to -u_I$ which is
an element of the ${\bf Z}_{2k}$ that we already identified. Thus, after imposing all
the identifications, the moduli space
is precisely ${\bf C}^4/\widehat{\bf D}_k$.

\subsec{The moduli space of the residual $USp(2l)_k \times
O(1)_{-2k}$ theories}

In these theories we can think of the massless matter fields simply
as fundamentals of $USp(2l)$, since the $O(1)$ gauge group
is just a discrete gauged ${\bf Z}_2$. It is enough to analyze
the case of $l=1$, since the analysis for general $l$ will just be $l$
copies of this.

The F-term equations imply that $A_1^T A_2 J A_2^T = A_2^T
A_2 J A_1^T$, but since $A_2$ is just a vector we have that
$A_2 J A_2^T = 0$ for any $A_2$. Thus $A_2 J
A_1^T = 0$ (or $A_2=0$), and similarly $A_1 J A_2^T=0$ (or $A_1=0$).
We conclude that $A_1$ and $A_2$ must be proportional to each other,
$A_2 = \alpha A_1$ for some $\alpha \in {\bf C} $.

It is obvious that $\sigma_{O(1)} = 0$, since there is no $O(1)$ Lie algebra,
while $\sigma_{USp}$ is equal to $\sigma_{USp} = A_1^{\dag} A_1 + J A_1^T
A_1^* J + A_2^{\dag} A_2 + J A_2^T A_2^* J$. The
constraint from the $\sigma$-term of the first field
is then
\eqn\spdterm{0 = A_1 \sigma_{USp} = A_1 A_1^{\dag} A_1 + A_1
A_2^{\dag} A_2 = (1+|\alpha|^2) (A_1 A_1^{\dag}) A_1,}
since $A_2$ is proportional to $A_1$, and $A_1 J A_1^T = 0$. This is
impossible unless $A_1=A_2=0$, since $(1+|\alpha|^2) (A_1
A_1^{\dag})$ is positive. Thus, the only classical solution to
all the
constraints is $A_1=A_2=0$.

We conclude that there is no moduli space associated to this
residual theory, in spite of the presence of massless fundamentals
at its unique classical vacuum.

\listrefs

\bye